\documentclass[%
 reprint,
 noeprint,  %
superscriptaddress,
 amsmath,amssymb,
 aps,
]{revtex4-2}

\usepackage{graphicx}%
\usepackage{dcolumn}%
\usepackage{bm}%

\usepackage{siunitx}
\DeclareSIUnit\ions{ions}
\DeclareSIUnit\ipsn{\ions \per \nm\squared}
\usepackage[version=4]{mhchem}
\usepackage{booktabs}

\usepackage[disable]{todonotes}

\usepackage{xcolor}
\definecolor{lightblue}{cmyk}{0.230667, 0.076, 0., 0.0706667}	%
\definecolor{darkred}{rgb}{0.717647,   0.086275,   0.086275}	%
\definecolor{darkred}{rgb}{0.85882   0.10196   0.10196}	%

\usepackage{hyperref}
\hypersetup{
    unicode=false,
    pdftoolbar=true,
    pdfmenubar=true,
    pdffitwindow=false,
    pdfstartview={FitH},
    pdftitle={Site-Selective Enhancement of Superconducting Nanowire Single-Photon Detectors via Local Helium Ion Irradiation},
    pdfauthor={Stefan Strohauer},
    pdfcreator={Latex with hyperref},
    pdfproducer={pdflatex},
    pdfkeywords={SSPD} {SNSPD} {He ion irradiation} {superconducting thin film} {transport measurement} {radiation damage} {He ion irradiation} {SNSPD array} {high-thickness SNSPD} {defect engineering},
    pdfnewwindow=true,
    pdfborder={0 0 0.5},
    colorlinks=false,
    linkcolor=red,
    citecolor=green,
    filecolor=magenta,
    urlcolor=cyan,
    allbordercolors={lightblue},
}
\usepackage[capitalize,noabbrev]{cleveref}

\newcommand{\dif}{\mathop{}\!\mathrm{d}}
\makeatletter
\newcommand{\spx}[1]{%
  \if\relax\detokenize{#1}\relax
    \expandafter\@gobble
  \else
    \expandafter\@firstofone
  \fi
  {^{#1}}%
}
\makeatother

\newcommand{\genericdel}[4]{%
  \ifcase#3\relax
  \ifx#1.\else#1\fi#4\ifx#2.\else#2\fi\or
  \bigl#1#4\bigr#2\or
  \Bigl#1#4\Bigr#2\or
  \biggl#1#4\biggr#2\or
  \Biggl#1#4\Biggr#2\else
  \left#1#4\right#2\fi
}

\makeatletter
\newcommand\thefontsize{The current font size is: \f@size pt}

\usepackage{xspace}	%

\newcommand{\sspd}{SNSPD\xspace}
\newcommand{\sspds}{SNSPDs\xspace}

\newcommand{\cls}{CLs\xspace}
\newcommand{\nbtin}{\ce{NbTiN}\xspace}
\newcommand{\nbn}{\ce{NbN}\xspace}
\newcommand{\Isw}{I_\mathrm{sw}\xspace}
\newcommand{\jsw}{j_\mathrm{sw}\xspace}
\newcommand{\Rsheet}{R_\mathrm{sheet}\xspace}
\newcommand{\Tc}{T_\mathrm{c}\xspace}
\newcommand{\electrondensity}{n_\mathrm{e}\xspace}
\newcommand{\tdecay}{\tau_\mathrm{d}\xspace}

\begin{document}

\title{Site-Selective Enhancement of Superconducting Nanowire Single-Photon Detectors via Local Helium Ion Irradiation}

\author{Stefan Strohauer}
\email{stefan.strohauer@wsi.tum.de}
\affiliation{Walter Schottky Institute, Technical University of Munich, 
85748 Garching, Germany
}
\affiliation{TUM School of Natural Sciences, Technical University of Munich, 
85748 Garching, Germany}
\author{Fabian Wietschorke}
\author{Lucio Zugliani}
\author{Rasmus Flaschmann}
\author{Christian Schmid}
\affiliation{Walter Schottky Institute, Technical University of Munich, 
85748 Garching, Germany
}
\affiliation{TUM School of Computation, Information and Technology, Technical University of Munich, 
80333 Munich, Germany
}
\author{Stefanie Grotowski}
\affiliation{Walter Schottky Institute, Technical University of Munich, 
85748 Garching, Germany
}
\affiliation{TUM School of Natural Sciences, Technical University of Munich, 
85748 Garching, Germany}
\author{Manuel Müller}
\affiliation{Walther-Meißer-Institut, 
85748 Garching, Germany
}
\affiliation{TUM School of Natural Sciences, Technical University of Munich, 
85748 Garching, Germany}
\author{Björn Jonas}
\affiliation{Walter Schottky Institute, Technical University of Munich, 
85748 Garching, Germany
}
\affiliation{TUM School of Computation, Information and Technology, Technical University of Munich, 
80333 Munich, Germany
}
\author{Matthias Althammer}
\affiliation{Walther-Meißer-Institut, 
85748 Garching, Germany
}
\affiliation{TUM School of Natural Sciences, Technical University of Munich, 
85748 Garching, Germany}
\author{Rudolf Gross}
\affiliation{Walther-Meißer-Institut, 
85748 Garching, Germany
}
\affiliation{TUM School of Natural Sciences, Technical University of Munich, 
85748 Garching, Germany}
\affiliation{Munich Center for Quantum Science and Technology (MCQST), 
80799 Munich, Germany}
\author{Kai Müller}
\affiliation{Walter Schottky Institute, Technical University of Munich, 
85748 Garching, Germany
}
\affiliation{TUM School of Computation, Information and Technology, Technical University of Munich, 
80333 Munich, Germany
}
\affiliation{Munich Center for Quantum Science and Technology (MCQST), 
80799 Munich, Germany}
\author{Jonathan Finley}
\email{finley@wsi.tum.de}
\affiliation{Walter Schottky Institute, Technical University of Munich, 
85748 Garching, Germany
}
\affiliation{TUM School of Natural Sciences, Technical University of Munich, 
85748 Garching, Germany}
\affiliation{Munich Center for Quantum Science and Technology (MCQST), 
80799 Munich, Germany}

\date{May 23, 2023}

\begin{abstract}

Achieving homogeneous performance metrics between nominally identical pixels is challenging for the operation of arrays of superconducting nanowire single-photon detectors (\sspds).
Here, we utilize local helium ion irradiation to post-process and tune single-photon detection efficiency, switching current, and critical temperature of individual devices on the same chip.
For \qty{12}{\nm} thick highly absorptive \sspds, which are barely single-photon sensitive prior to irradiation, we observe an increase of the system detection efficiency from $<\qty{0.05}{\percent}$ to \qty{55.3+-1.1}{\percent} following irradiation.
Moreover, the internal detection efficiency saturates at a temperature of \qty{4.5}{\K} after irradiation with \qty{1800}{\ipsn}.
For irradiated \qty{10}{\nm} thick detectors we observe a doubling of the switching current (to \qty{20}{\uA}) compared to \qty{8}{\nm} \sspds of similar detection efficiency, increasing the amplitude of detection voltage pulses.
Investigations of the scaling of superconducting thin film properties with irradiation up to a fluence of \qty{2600}{\ipsn} revealed an increase of sheet resistance and a decrease of critical temperature towards high fluences.
A physical model accounting for defect generation and sputtering during helium ion irradiation is presented and shows good qualitative agreement with experiments. 
\end{abstract}


\maketitle

\section{Introduction}

Superconducting Nanowire Single-Photon Detectors (\sspds) \cite{Goltsman2001}  play a significant role in quantum technologies \cite{Takesue2007, 
Chen2022, Liu2023, Bussieres2014, McCarthy2013, Grein2015, Takesue2015, Valivarthi2016}
and a wide range of applications requiring general faint light detection \cite{Zhang2003, Tanner2011}. 
Compared to Single-Photon Avalanche Diodes (SPADs) \cite{Itzler2011}, their superior performance metrics, consisting of high detection efficiency also at long wavelengths \cite{Marsili2012a, Korneev2012}, 
low dark count rate \cite{Shibata2015}, and low timing jitter \cite{Korzh2020} make them ideally suited for demanding applications such as quantum key distribution \cite{Takesue2007, 
Chen2022, Liu2023}, quantum computing \cite{Natarajan2012}, or deep space optical communication \cite{Grein2015}.
Moreover, their waveguide-integrated form is a key component for photonic integrated circuits \cite{Ferrari2018, Sprengers2011, Reithmaier2013,
Reithmaier2015, Majety2023}.

Since recently, \sspds also find application in fields such as astronomy \cite{Wollman2021}, dark matter detection \cite{Chiles2022}, and particle detection \cite{Polakovic2020, Shigefuji2023}.
However, these applications typically require large detector arrays or even an \sspd camera, which to date turns out to be challenging due to the necessary readout and homogeneity within an ensemble of the order of hundreds to thousands of detectors.
Recently, row-column multiplexing of a 1024-pixel array \cite{Wollman2019}, and a promising readout architecture based on thermal coupling and time-of-flight measurements \cite{McCaughan2022} were demonstrated.
For such pixel arrays, typically amorphous materials such as \ce{MoSi} and \ce{WSi} are used, although \sspds based on polycrystalline materials like \nbn and \nbtin exhibit higher critical temperatures, larger critical currents, and lower timing jitter.
Compared to polycrystalline materials and their spatial inhomogeneities of the superconducting energy gap \cite{Hortensius2013, Noat2013, Sacepe2008, Kirtley1987}, amorphous films attain better homogeneity and the associated higher yield of similarly performing detectors \cite{Allman2015, Gaudio2014, Kerman2007, EsmaeilZadeh2021, Marsili2013}.
To enable the use of \nbn for large pixel arrays, atomic layer deposition and molecular-beam epitaxy of highly homogeneous films have been investigated recently as alternatives to the common deposition of polycrystalline \nbn and \nbtin films grown using reactive magnetron sputtering \cite{Steinhauer2021a, Cheng2019a, Knehr2019, Cheng2020}.
In addition to methods for obtaining better homogeneity during film deposition, a method to tune detector metrics of individual devices after fabrication would also be highly advantageous.
Inspired by the recent work of 
\textcite{Zhang2019}, 
which sparked interest in irradiating \sspds with helium (He) ions \cite{Xu2021, Zhang2021, Zhang2022}, we use a He ion microscope as a post-processing tool to tune detector metrics of individual \nbtin devices fabricated on the same chip.
At the same time, we investigate how \sspd properties such as detection efficiency and switching current depend on the He ion fluence.
In addition to detector metrics, we explore the scaling of \nbtin thin film parameters such as sheet resistance and critical temperature with increasing irradiation.

\section{Experimental}
\label{sec:experimental}
To study the influence of He ion irradiation on the native transport properties of \nbtin thin films and the performance of \sspds, we deposited \nbtin films with thicknesses of \qty{8}{\nm}, \qty{10}{\nm}, and \qty{12}{\nm} using DC reactive magnetron sputtering onto \ce{Si} substrates with a \qty{130}{\nm} thick thermally grown \ce{SiO2} layer.
The \nbtin thickness was controlled by measuring the sputtering rate and adjusting the sputtering time correspondingly.
Subsequently, we patterned the \nbtin films into cloverleaf structures and \sspds using electron beam lithography and reactive ion etching, followed by optical lithography and gold evaporation for contact pad fabrication \cite{Flaschmann2023}.
The detector design consists of a \qty{100}{\nm} wide wire in a meander form with a fill factor of \qty{50}{\percent}, and a total active area of \qtyproduct[product-units = repeat]{10x10}{\um}.
The cloverleaf structures were fabricated in order to perform magneto-transport measurements in van-der-Pauw geometry \cite{vanderPauw1958,Miccoli2015} with an active area of \qtyproduct[product-units=repeat]{10x10}{\um}
and to correlate the results of macroscopic transport with the He ion fluence dependent performance metrics of the corresponding \sspds.
To ensure the best comparability, cloverleafs (\cls) and \sspds were fabricated on the same chip.
For this study, they were subsequently irradiated with a He ion microscope (Zeiss Orion Nanofab) with He ion fluences ranging from \qty{0}{\ipsn} to \qty{2600}{\ipsn}.

The magneto-transport measurements were performed by cooling the samples to \qty{4.2}{\K} before allowing them to slowly heat up to \qty{20}{\K} in external magnetic fields between \qty{-0.1}{\tesla} and \qty{1}{\tesla}, applied perpendicular to the sample plane. 
From these measurements, we extract the sheet resistance of the superconducting thin film at \qty{20}{\K} and room temperature, the critical temperature of the superconducting thin film, and the Bogoliubov quasiparticle diffusivity.
Also, by measuring the \cls in Hall geometry and performing magnetic field sweeps, followed by a linear fit of the Hall voltage, we determine the Hall coefficient and electron density of the \nbtin films.\cite{Sidorova2021}

Switching current $\Isw$ and system detection efficiency (SDE) of the \sspds were measured using a cryogenic probe station (Janis) at \qty{4.5}{\K}.
To calculate the SDE, we determined the dark count rate (DCR) before we measured the count rate (CR) by homogeneous illumination of the \sspd with an attenuated \qty{780}{\nm} continuous wave diode laser and polarization parallel to the nanowire.
The SDE is then defined as $\mathrm{SDE} = \frac{\mathrm{CR}-\mathrm{DCR}}{\mathrm{PR}}$ with the photon rate PR incident on the cryogenic probe station.

\section{Results and discussion}
In this section, we present the dependence of \nbtin thin film properties and detector metrics on He ion irradiation for film thicknesses of \qty{8}{\nm}, \qty{10}{\nm}, and \qty{12}{\nm}.
Provided that the \sspds are sensitive to single photons, using larger thicknesses for \sspds generally results in stronger optical absorption \cite{Semenov2009,Banerjee2018} and therefore enhances their overall system detection efficiency (SDE).
Moreover, we aim for a better understanding of how He ion irradiation modifies the transport properties of the \nbtin film and focus on establishing structure-property relationships that link detector thickness, He ion fluence, and detector performance.

\subsection{Performance of He ion irradiated \sspds}
\begin{figure}
 \centering
 \includegraphics{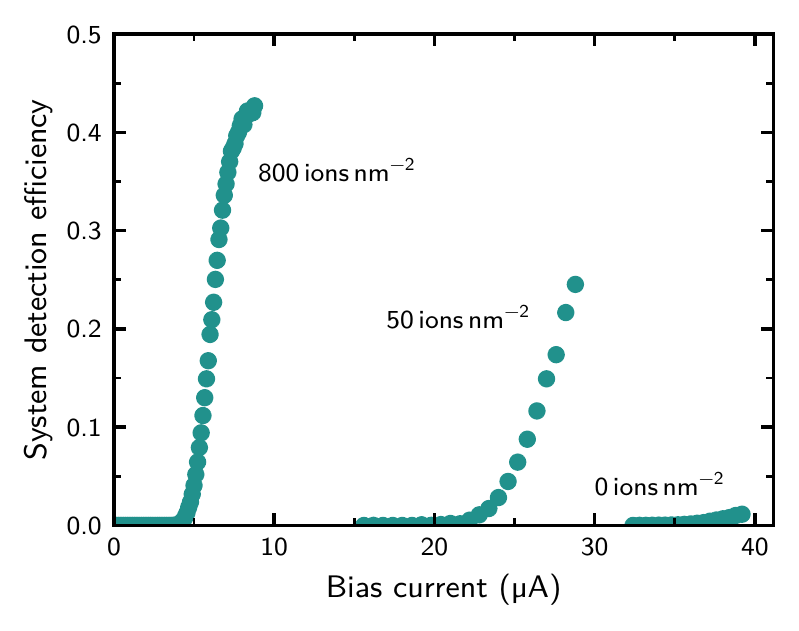}
     \caption{System detection efficiency vs. bias current of the same \qty{10}{\nm} thick detector for three different He ion fluences. 
     The relative uncertainties of He ion fluence, SDE, and bias current are \qty{5}{\percent}, \qty{2}{\percent} and less than \qty{1}{\percent}, respectively (error bars not shown for clarity).
     With increasing fluence the efficiency rises up to $43\%$ and shows the beginning of saturating internal detection efficiency, while the switching current decreases. The largest change in SDE and $\Isw$ is induced by the first He ions that hit the detector.}
 \label{fig:efficiency_at_three_doses}
\end{figure}

\Cref{fig:efficiency_at_three_doses} shows the increase in SDE and the simultaneous decrease of switching current of a representative \qty{10}{\nm} thick device measured before irradiation and after fluences of \qty{50}{\ipsn} and \qty{800}{\ipsn}.
Irradiating the detector with \qty{50}{\ipsn} already results in an increase in SDE from $< \qty{2}{\percent}$ to \qty{25}{\percent}.
At a fluence of \qty{800}{\ipsn} the detector shows the beginning of saturating SDE at \qty{43}{\percent}, close to the maximum absorption of \qty{53.1}{\percent} in the detector as determined by finite-difference time-domain (FDTD) simulations (\cref{app:absorption-simulation}).
Simultaneously, a decrease in switching current $\Isw$, which is defined as the maximum current the detector can sustain before switching to the normal conducting state, is apparent and ranges from \qty{39.2}{\uA} to \qty{28.8}{\uA} and \qty{8.6}{\uA} after irradiation.

\begin{figure}
 \centering
 \includegraphics{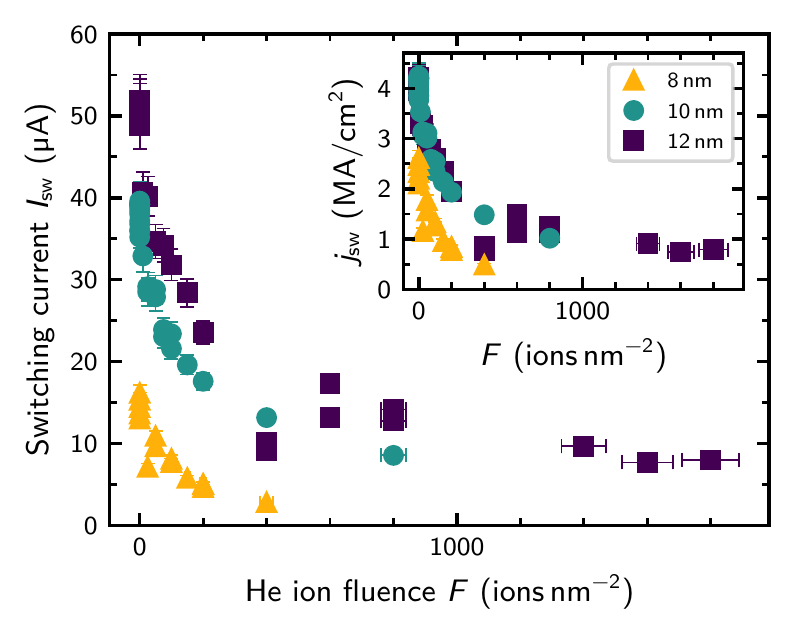}
 \caption{Switching current vs. He ion fluence for \qty{8}{\nm}, \qty{10}{\nm}, and \qty{12}{\nm} detector thickness, including statistical errors.
 $\Isw$ decreases with the He ion fluence, showing the largest decrease for low fluences.
 A strong dependence of $\Isw$ on the film thickness is apparent throughout the whole fluence range studied. 
 The inset shows the switching current density $\jsw$ as calculated from $\Isw$ and the wire width and thickness, accounting for an effective thickness reduction due to surface sputtering during He ion irradiation as well as a \qty{1.3}{\nm} thick native NbTiN oxide.
 }
 \label{fig:I_sw_vs_dose}
\end{figure}

To study the scaling of the switching current with He ion fluence, we irradiated multiple detectors that have thicknesses of \qty{8}{\nm}, \qty{10}{\nm}, and \qty{12}{\nm} using different He ion fluence values.
\Cref{fig:I_sw_vs_dose} shows the resulting data, revealing a clear trend of decreasing $\Isw$ with He ion fluence.
As expected, $\Isw$ is higher for thicker devices of the same fluence due to the larger cross-sectional area of thicker nanowires.
We explain the scattering of measured switching currents of nominally identical detectors by constrictions that limit $\Isw$ to a lower value than non-constricted devices have.\cite{Kerman2007}
Such scatter is particularly visible by the large variation of currents of the non-irradiated devices and the small values of the two \qty{12}{\nm} detectors irradiated with \qty{400}{\ipsn}.
The inset of \cref{fig:I_sw_vs_dose} shows the switching current density $\jsw$ as calculated from $\Isw$ and the cross sectional area of the wire, given by the width and the nominal wire thickness presented in \cref{tab:absorption}.
Furthermore, we accounted for an effective reduction of the nominal thickness due to surface sputtering during He ion irradiation as derived in \cref{sec:scaling-thin-film-with-fluence} and for a native \nbtin oxide of \qty{1.3}{\nm} thickness \cite{Zhang2018b}.
The switching current density of our non-irradiated \nbtin detectors is comparable to the data \textcite{Korneeva2018} present for \qty{5.8}{\nm} thick \ce{NbN} devices.
Moreover, we observe $\jsw$ of the \qty{8}{\nm} film being smaller than for the \qty{10}{\nm} and \qty{12}{\nm} films.
We note that for thin and narrow wires, the depairing current density \cite{Clem2012, Korneeva2018} can limit the measurable switching current density and reveals a dependence on the film thickness \cite{Ilin2008,Ilin2005}.
Thus, an increased depairing current density for the thicker devices also likely contributes to their higher  $\Isw$ and $\jsw$.
\begin{figure}
 \centering
 \includegraphics{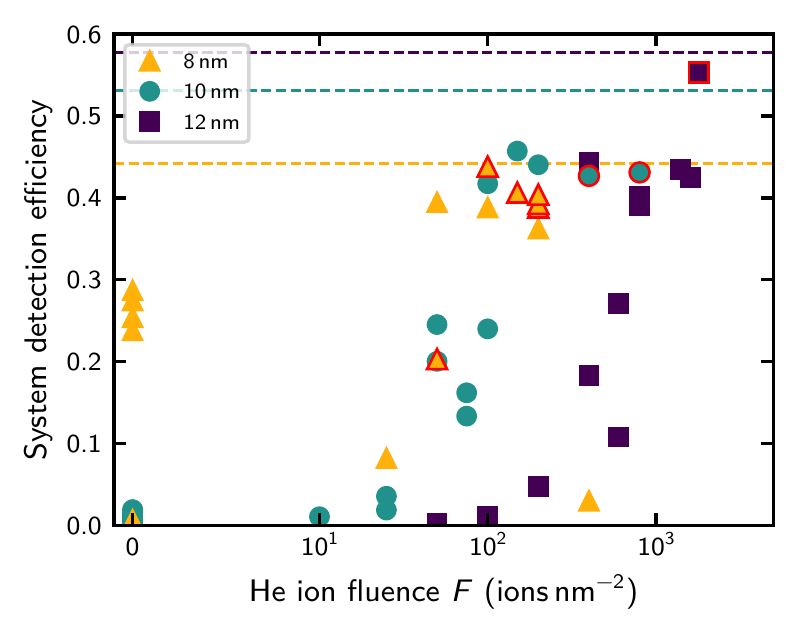}
 \caption{System detection efficiency vs. He ion fluence for the three detector thicknesses studied in this work. Dashed lines indicate the absorption in the \sspd simulated with FDTD; data points with saturating SDE are highlighted with a red frame.
 The relative uncertainties of SDE and He ion fluence are \qty{2}{\percent} and \qty{5}{\percent}, respectively (error bars not shown for clarity).
 Each of the data points stems from a different detector that was irradiated once with the given dose except for two \qty{10}{\nm} and two \qty{12}{\nm} detectors that were irradiated twice; for some \sspds we measured the SDE in addition also before irradiation.
 Despite the large scattering of data points that can be explained by the strong variation of the initial SDE between individual devices, one clearly sees that the SDE increases with He ion fluence and that the total maximum SDE is reached by the largest detector thickness.
 }
 \label{fig:DE_max_vs_dose_log}
\end{figure}

Similar to $\Isw$, we investigated the scaling of SDE with the He ion fluence.
As shown in \cref{fig:DE_max_vs_dose_log}, we observe an increase of SDE with the He ion fluence for all detector thicknesses, despite the large scatter between data obtained from nominally identical \sspds.
Most notably, the SDE for the \qty{12}{\nm} thick detectors increases from less than \qty{0.05}{\percent} for the non-irradiated case to \qty{55.3}{\percent} and just saturating detection efficiency for a fluence of \qty{1800}{\ipsn}.
As expected, the SDE increases with detector thickness due to the higher absorption.
The dashed horizontal lines in \cref{fig:DE_max_vs_dose_log} show the upper limit for the SDE, defined by the absorption of \sspds of the respective thicknesses that we obtained from FDTD simulations discussed in \cref{app:absorption-simulation}.
We note that one can further enhance the absorption and thus the SDE over a broad wavelength range by adding a metal mirror with an optical cavity underneath the \sspd.\cite{Li2019b}
Recently, a similar approach for He ion irradiated detectors involving a narrow-band cavity, realized with a distributed Bragg reflector, has shown to push the absorption to over \qty{90}{\percent}.\cite{Xu2021}
The fact that the measured SDE of the highest irradiated \qty{8}{\nm} detector shown in \cref{fig:DE_max_vs_dose_log} is less than \qty{3}{\percent} can be explained as follows:
Due to the irradiation-induced reduction of the switching current (to \qty{2.8}{\uA} for this detector), which is also the maximum applicable bias current, the maximum voltage pulse amplitude decreases as well.
However, the trigger level of the counter used to measure the efficiency can only be reduced correspondingly as long as it is well above the electrical noise floor. This implies that once the pulse amplitude becomes comparable to the electrical noise floor, a substantial fraction of detection pulses will not be registered by the counter anymore.
Depending on the readout electronics used and on their noise floor, this sets the limit for meaningful He ion fluences when irradiating \sspds.
Surprisingly, one of the \qty{12}{\nm}/\qty{400}{\ipsn} detectors shown in \cref{fig:DE_max_vs_dose_log} exhibits a high SDE although $\Isw$ was lower than expected for these two \sspds.
This hints to a relatively homogeneous current density within the nanowire that allows biasing close to the depairing current density and thus achieving high SDE.

\begin{figure}
 \centering
 \includegraphics{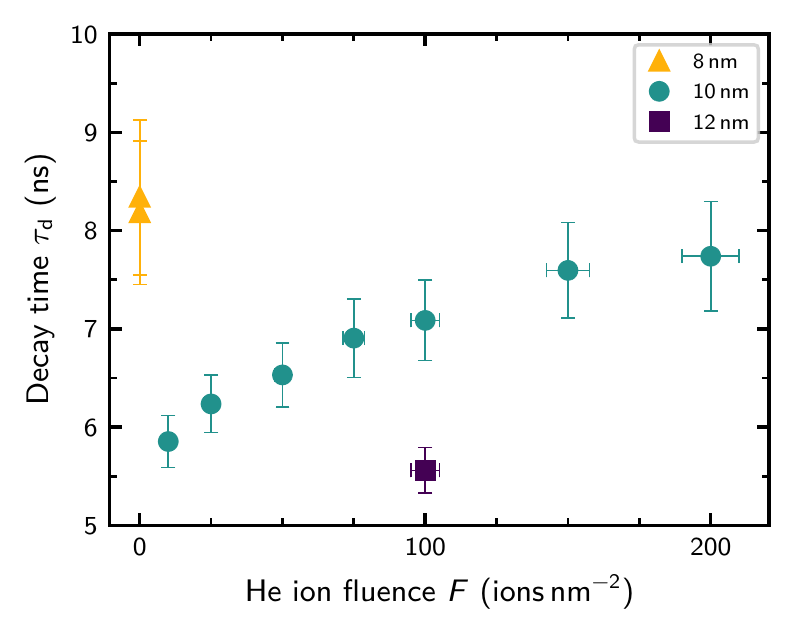}
 \caption{Decay time vs. He ion fluence, including statistical errors.  $\tdecay$ increases with increasing fluence and decreasing thickness due to the resulting higher kinetic inductance of the detector.
}
 \label{fig:fall_time}
\end{figure}

Another key metric for \sspds is their recovery time since it determines the detector's maximum count rate.
It can be estimated from the time constant $\tdecay$ of the exponential decay of a detection voltage pulse.\cite{Kerman2006,Ferrari2018}
\Cref{fig:fall_time} shows how the measured decay time increases with increasing He ion fluence and demonstrates that it is smaller for thicker detectors.
These observations can be understood as follows:
The decay time depends on the kinetic inductance $L_\mathrm{k}$ of the detector by $\tdecay = L_\mathrm{k}/R_\mathrm{load}$ with a typical load resistance of $R_\mathrm{load} = \qty{50}{\ohm}$ for the readout electronics.\cite{Kerman2006}
At the same time, $L_\mathrm{k}$ for a thin ($d \ll \lambda_\mathrm{eff}$) and dirty ($\ell \ll \xi_0$) film  of length $l$, width $w$, and thickness $d$ is given by
\begin{equation}
    L_\mathrm{k}
    =
    \mu_0 \lambda_\mathrm{eff, tf} \, \frac{l}{w} \;,
    \label{eq:kinetic-inductance-fundamental}
\end{equation}
with the effective magnetic penetration depth for thin films $\lambda_\mathrm{eff, tf} = \lambda_\mathrm{eff}^2/d$ as introduced by \textcite{Pearl1964}, where $\lambda_\mathrm{eff}$ is the effective magnetic penetration depth of a dirty bulk superconductor like \nbtin, given by
\begin{equation}
    \lambda_\mathrm{eff}
    =
    \lambda_\mathrm{L}
    \sqrt{\frac{\xi_0}{\ell}}
    =
    \sqrt{\frac{\hbar \rho}{\pi \mu_0 \Delta(\qty{0}{\K})}} %
    \label{eq:effective-magnetic-penetration-depth}
\end{equation}
according to \textcite[][Eq.~(9.36)]{Bartolf2016}.
Here, $\lambda_\mathrm{L}$ is the London penetration depth, $\xi_0$ the BCS coherence length, $\ell$ the mean free path, $\rho$ the specific resistivity of the superconducting film in the normal conducting state, and $\Delta(\qty{0}{\K})$ the superconducting energy gap.\footnote{
For a detailed derivation of \cref{eq:kinetic-inductance-fundamental,eq:effective-magnetic-penetration-depth} 
see also the book of \textcite[][Eqs.~(3.120), (3.137), and (6.67b)]{Tinkham2004}.
}
Hence, with the effective magnetic penetration depth one can express the kinetic inductance as
\begin{equation}
    L_\mathrm{k}
    =
    \mu_0 \frac{\lambda_\mathrm{eff}^2}{d}\frac{l}{w}
    =
    \frac{\hbar \Rsheet}{\pi \Delta(\qty{0}{\K})}
    \frac{l}{w}
    \;.
    \label{eq:kinetic-inductance-practical}
\end{equation}
Thus, for detectors of similar length and width, the kinetic inductance and the decay time are smaller for detectors that exhibit a smaller sheet resistance, for example due to the use of a thicker film or due to less irradiation with He ions.
In this way, we conclude that the increase of decay time due to irradiation can be compensated to a certain extent by using thicker films.

\begin{figure}
 \centering
 \includegraphics{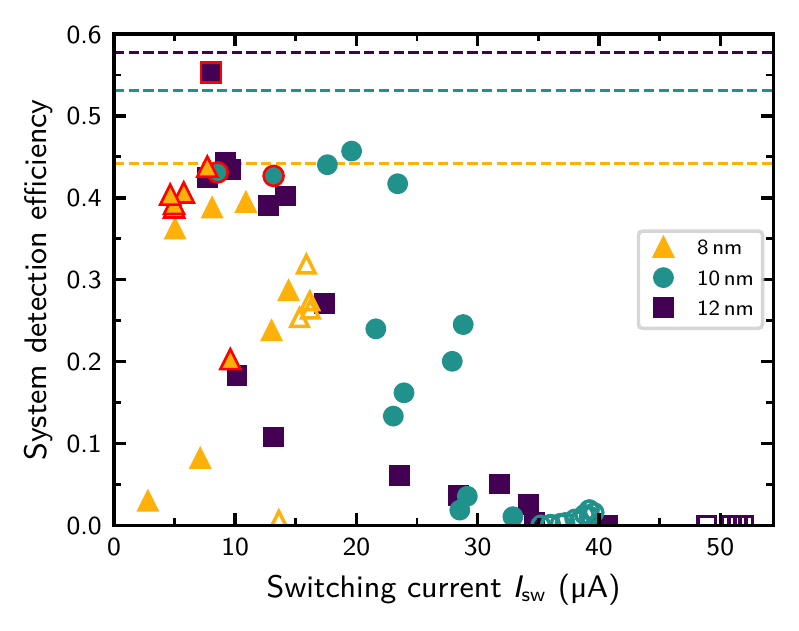}
 \caption{Maximum system detection efficiency vs. switching current for the three detector thicknesses studied in this work. The simulated maximum SDE of each thickness is indicated by dashed lines, while saturating SDE is highlighted by symbols with a red frame; open symbols represent non-irradiated devices.
 The relative uncertainties of SDE and switching current are \qty{2}{\percent} and \qty{6}{\percent}, respectively (error bars not shown for clarity).
 Noteworthy are the data points at an $\mathrm{SDE} \approx \qty{45}{\percent}$, where the \qty{10}{\nm} \sspds provide a similar SDE like the \qty{8}{\nm} ones but offer the doubled switching current.
 Furthermore, for a similar $\Isw$ of about \qty{8}{\uA}, one \qty{12}{\nm} \sspd shows up to \qty{55.3}{\percent} SDE, whereas the \qty{8}{\nm} ones provide only up to \qty{43.8}{\percent} SDE.
}
 \label{fig:DE_vs_I_sw}
\end{figure}

For applications, simultaneously high SDE and $\Isw$ are desired since a higher $\Isw$ yields a higher detection pulse, which reduces not only the requirements for pulse detection with the readout electronics but also the timing jitter induced by electrical noise \cite{Korzh2020, Flaschmann2023}.
To compare these two performance metrics, \cref{fig:DE_vs_I_sw} shows the SDE against $\Isw$ (open symbols representing the non-irradiated detectors, a red frame highlighting saturating SDE, and dashed lines indicating the simulated SDE upper limit).
It is interesting to note how $\Isw$ and SDE compare between the \qty{8}{\nm} and the \qty{10}{\nm} devices with an SDE between \qty{39}{\percent} and \qty{46}{\percent}: While providing a similar efficiency, the \qty{10}{\nm} devices offer twice as much switching current, \qty{20}{\uA} instead of \qty{10}{\uA}.
This $\Isw$ is also higher than that of the non-irradiated \qty{8}{\nm} detectors.
Another comparison can be drawn between the \qty{8}{\nm} \sspds with saturating SDE close to \qty{44}{\percent} and the \qty{12}{\nm} \sspd showing \qty{55.3}{\percent} SDE: at similar switching currents of about \qty{8}{\uA}, the \qty{12}{\nm} \sspd provides a substantially higher SDE.
Furthermore, it is noteworthy that the shift of the data point clouds in this two-dimensional parameter space is not monotonous to higher $\Isw$ with higher thickness (except for the non-irradiated devices), which could hint to the existence of an optimum thickness between \qty{8}{\nm} and \qty{12}{\nm} to reach simultaneous high SDE and $\Isw$ via He ion irradiation.

To conclude, by choosing a suitable detector thickness and He ion fluence, one can tune $\Isw$ and SDE, even to better performance in both parameters simultaneously compared to non-irradiated detectors.
Moreover, by individual irradiation of \nbtin \sspds with a suitable He ion fluence, one can intentionally modify the performance of selected detectors or even mitigate differences between nominally identical devices.

\subsection{Scaling of thin film metrics with He ion fluence}
\label{sec:scaling-thin-film-with-fluence}
\begin{figure}
 \centering
 \includegraphics{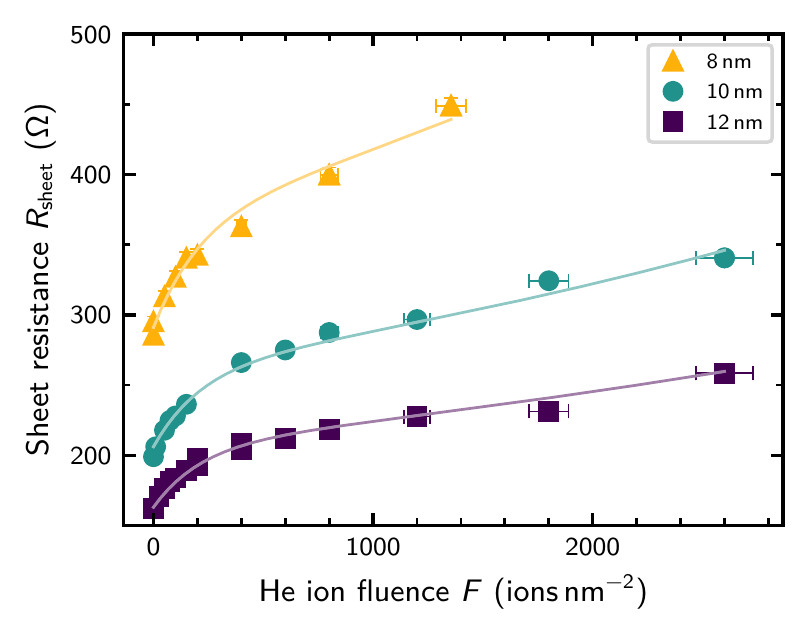}
 \caption{Sheet resistance vs. He ion fluence for \qty{8}{\nm}, \qty{10}{\nm}, and \qty{12}{\nm} film thickness, including statistical errors. 
 The sheet resistance increases with He ion fluence and decreasing film thickness. 
 All three data sets are described by the fit function given by \cref{eq:R_sheet_phyiscal_fit} with the parameters of \cref{tab:fit-parameters}.
 } 
 \label{fig:R_sheet_vs_dose}
\end{figure}

To investigate how He ion irradiation impacts upon the bare \nbtin film metrics such as critical temperature $\Tc$, sheet resistance $\Rsheet$, and electron density $\electrondensity$, we fabricated cloverleaf structures together with the detectors on the same sample to perform magneto-transport measurements in van-der-Pauw geometry.
In \cref{fig:R_sheet_vs_dose} we present the dependence of the sheet resistance $\Rsheet$ on the He ion fluence. 
As expected, $\Rsheet$ is higher for thinner films and increases with increasing He ion fluence as the number of defects in the \nbtin film increases.
Interestingly, the sheet resistance does not scale as $\Rsheet = \rho/d_0$ with the nominal film thickness $d_0$ as expected if all samples had the same specific resistivity $\rho$.
Even if one subtracts a \qty{1.3}{\nm} thick layer of oxidized \nbtin \cite{Zhang2018b} from the nominal \nbtin thickness, the resulting resistivities of the non-irradiated films are still lower for the thicker films than for the \qty{8}{\nm} film (\qty{1.94}{\micro\ohm\m}, \qty{1.73}{\micro\ohm\m}, and \qty{1.73}{\ohm\m} for the \qty{8}{\nm}, \qty{10}{\nm}, and \qty{12}{\nm} films, respectively).
Although one might expect $\Rsheet$ to saturate at high fluences due to a saturating defect density in the film, we experimentally observe a continuous increase of $\Rsheet$ with He ion fluence.
This could have its origin in noticeable surface sputtering \cite{Alkemade2012} and intermixing \cite{Li2009} at the film/substrate interface by the impinging He ions and an associated reduction of the effective film thickness.

Based on these observations and taking the sheet resistance to be directly proportional to the defect density, we develop a simple physical model.
In our model, each ion that passes through the film can create a defect cluster of an average volume $v_\mathrm{D}$ with an efficiency $\eta$.
Moreover, we consider the film volume $V$ as divided into many volume elements with the same size as the average defect cluster volume $v_\mathrm{D}$, and defect clusters may only be created in volume elements that do not already contain a defect cluster.
Those considerations imply that irradiating a film of volume $V$, thickness $d$, and area $A$ using a He ion fluence $\Delta F$ creates $\Delta N_\mathrm{D}$ new defect clusters according to
\begin{align}
        \Delta N_\mathrm{D}
    &=  \left(
        \frac{V-N_\mathrm{D} \, v_\mathrm{D}}{V}
        \right)
        \left(
        \frac{d}{\sqrt[3]{v_\mathrm{D}}}
        \right)
        \eta \, A \, \Delta F
        \;.
\end{align}
The first fraction represents the fraction of $V$ that does not yet contain defect clusters, the second fraction represents the number of potential defect clusters that an impinging ion could create when passing the film along its thickness.
Dividing this equation by the total volume $V$ to obtain an expression for the defect cluster density $n_\mathrm{D}$ and taking the limit $\Delta F \rightarrow 0$ yields
\begin{align}
        \frac{\dif n_\mathrm{D}}{\dif F}
    &=  \frac{\eta}{\sqrt[3]{v_\mathrm{D}}}
        \left(1-v_\mathrm{D}\, n_\mathrm{D}(F)\right)
        \;.
\end{align}
This differential equation has the solution
\begin{align}
    n_\mathrm{D}(F)
    &=
    \frac{1}{v_\mathrm{D}}
    \left(1-
    (1-n_{\mathrm{D},0}\, v_\mathrm{D}) \, 
    e^{-\eta v_\mathrm{D}^{2/3} F}
    \right)
    \;,
\end{align}
where $n_{\mathrm{D},0}$ is the defect cluster density of the non-irradiated film.
We relate the defect cluster density to the specific resistivity $\rho$ via direct proportionality with a film-thickness dependent constant $a_{d_0}$.
To arrive at a model for the sheet resistance, we further account for the previously mentioned surface sputtering due to He ion bombardment by including an effective reduction of the original film thickness $d_0$ with a sputtering rate $r_\mathrm{s}$ and conclude
\begin{align}
    \Rsheet(F)
    &=
    \frac{1}{v_\mathrm{D}}
    \left(1-
    (1-n_{\mathrm{D},0} \, v_\mathrm{D}) \, 
    e^{-\eta v_\mathrm{D}^{2/3} F}
    \right)
    \frac{a_{d_0}}{d_0 - r_\mathrm{s} F}
    \;.
    \label{eq:R_sheet_phyiscal_fit}
\end{align}
We fit this model to the experimental data and present the results of this fitting in \cref{fig:R_sheet_vs_dose}.
Since the sputtering rate $r_\mathrm{s}$ and the factors $n_{\mathrm{D},0} \, v_\mathrm{D}$ and $\eta v_\mathrm{D}^{2/3}$  contain only thickness-independent quantities, we choose these factors as common fit parameters for all three thicknesses.
In this way, $a_{d_0}/v_\mathrm{D}$ is the only individual fit parameter for each thickness, while the other three previously mentioned parameters are shared between all films.
As such, we fit the three data sets with six parameters.
\Cref{tab:fit-parameters} lists the parameters that result in the fit functions shown in \cref{fig:R_sheet_vs_dose}.
\begin{table*}
	\centering
	\caption{Fit parameters of the physical model according to \cref{eq:R_sheet_phyiscal_fit}, describing the data in \cref{fig:R_sheet_vs_dose}. For each thickness, the fit function has its own  fit parameter $a_{d_0}/v_\mathrm{D}$, while the other parameters are independent of film thickness and therefore shared between the three fit functions.
 }
	\label{tab:fit-parameters}
\begin{tabular}{
S
!{\qquad}
S[table-format=4(2), table-align-exponent = false, table-align-uncertainty = false]
!{\qquad}
S[table-format=1.2(2), table-align-exponent = false, table-align-uncertainty = false]
!{\qquad}
S[table-format=1.1(1)e-1, table-align-exponent = false]
!{\qquad}
S[table-format=1.1(1)e-1, table-align-exponent = false]
}
    \toprule
    {$d_0$ (\unit{\nm})} & {$a_{d_0}/v_\mathrm{D}$ (\unit{\ohm})}   & {$n_{\mathrm{D},0} \, v_\mathrm{D}$ (\unit{1})} & {$\eta v_\mathrm{D}^{2/3}$ (\unit{\nm^2})} & {$r_\mathrm{s}$ (\unit{\nm\per(\ipsn)})}  \\
    \midrule
    8   & 2957(36)  & & & \\
    10  & 2618(36)  & 0.79(1)  & 4.7(7)e-3 & 9.4(6)e-4\\
    12  & 2484(32)  & & &  \\
    \bottomrule
\end{tabular}
\end{table*}
Considering the volume $V$ as divided into many volume elements with the same size as the average defect cluster volume $v_\mathrm{D}$, the parameter $n_{\mathrm{D},0} \, v_\mathrm{D}$ can be qualitatively interpreted as the fraction of volume elements of the non-irradiated film that contains transport-related defective regions such as grain boundaries and scattering centers.
We obtain $n_{\mathrm{D},0} \, v_\mathrm{D} = 0.79$, indicating that the initial defect cluster density is large and/or the average volume of a defect cluster induced by a single He ion collision is on the order of the \nbtin grain size.
A high defect density is not unexpected for a polycrystalline material such as \nbtin, and a defect cascade induced by a single He ion can extend over a volume similar to the \nbtin grain size (few \unit{\nm}) according to a study of He ion irradiation induced defect clusters in copper \cite{Shimizu1994}.
Furthermore, the quantity $\eta v_\mathrm{D}^{2/3}$ can be understood as the cross section determining the probability that an impinging He ion creates a defect cluster of volume $v_\mathrm{D}$.
Moreover, the sputtering rate of \qty{9.4e-4}{\nm\per(\ipsn)} implies that an irradiation by \qty{1000}{\ipsn} leads to an effective reduction of the \nbtin film thickness by 
about \qty{1}{\nm}.
Although \textcite{Zhang2019} did not observe a change in thickness after irradiating their \nbn film with \qty{500}{\ipsn}, our observation agrees well with the simulated and experimentally observed sputtering yield of typically \qty{1}{\nm} per \qty{1000}{\ipsn} found in literature \cite{Behrisch2007,Alkemade2012}.

\begin{figure}
 \centering
 \includegraphics{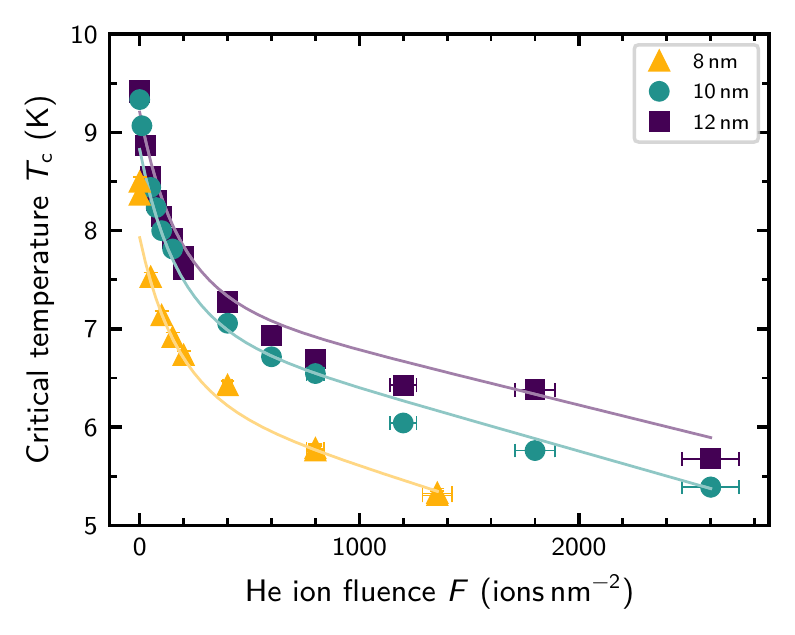}
 \caption{
 Critical temperature vs. He ion fluence, including statistical errors. $\Tc$ decreases with He ion fluence, with the reduction in $\Tc$ at small fluences being the strongest.
 Surprisingly, $\Tc$ for the unirradiated \qty{10}{\nm} and \qty{12}{\nm} films are similar for small fluences, although $\Tc$ is typically higher for thicker films.
 The continuous functions we determined by fitting the $\Tc$ and $\Rsheet$ data with the universal scaling law introduced by \textcite{Ivry2014}, $d_0 \Tc = A \Rsheet^{-B}$, and subsequently using our physical model for $\Rsheet$, given by \cref{eq:R_sheet_phyiscal_fit}, as input to the universal scaling law.
 }
 \label{fig:Tc_vs_dose}
\end{figure}

\Cref{fig:Tc_vs_dose} shows the dependence of the critical temperature $\Tc$ on the He ion fluence for \qty{8}{\nm}, \qty{10}{\nm}, and \qty{12}{\nm} thick films.  
Clearly, $\Tc$ decreases continuously by about \qty{30}{\percent} from the non-irradiated film to the film irradiated with \qty{1200}{\ipsn}.
Similarly to $\Rsheet$, also $\Tc$ decreases most strongly for small He ion fluences.
Interestingly, the measured values of $\Tc$ for the \qty{10}{\nm} and \qty{12}{\nm} films are very similar for low He ion fluences, although we would expect a lower $\Tc$ for the thinner film due to the suppression of superconducting properties when transitioning from bulk to the nanoscale.\cite{Holzman2019, Haviland1989, Bezryadin2000}
Furthermore, we fit our experimental data for $\Tc$ and $\Rsheet$ with the universal scaling law introduced by \textcite{Ivry2014}, $d_0 \Tc = A \Rsheet^{-B}$, which relates critical temperature, sheet resistance, and film thickness. 
Combining then the resulting fit function $\Tc(\Rsheet, d_0)$ with our physical fit function for $\Rsheet$, \cref{eq:R_sheet_phyiscal_fit}, we obtain the fits shown in \cref{fig:Tc_vs_dose}.
\Cref{app:Tc-fitting} contains details of the fitting procedure used for $\Tc$.
A recent publication by \textcite{Ruhtinas2023} contains a study of the critical temperature and the critical current density of comparably thick, \qty{35}{\nm} and \qty{100}{\nm}, \nbtin bridges, in which they suppressed superconductivity by He ion irradiation of a narrow line perpendicular to the bridge. 
Empirically, they observed a logarithmic dependence of $\Tc$ and an exponential dependence of $\jsw$ on the He ion fluence $F$.
For the critical temperature, a fit of our data with $\Tc(F) = -a\log(F+b)+c$ and fitting parameters $a$, $b$, and $c$ for each of the three thicknesses describes our data even a bit better than the universal scaling law.
However, using the universal scaling law, we need only the two fitting parameters $A$ and $B$ to describe all three data sets, while the empirical logarithmic fit function requires three fitting parameters for each thickness, a total of nine parameters for our three data sets.
Moreover, our data for $\jsw$ as shown in the inset of \cref{fig:I_sw_vs_dose} indicates that the switching current density does not follow the exponential dependence observed by \textcite{Ruhtinas2023}, especially for the smaller He ion fluences. 
However, we note that compared to our work, \textcite{Ruhtinas2023} studied the switching current density for higher fluences, ranging from \qty{2e4}{\ipsn} to \qty{12e4}{\ipsn}.
Furthermore, since the film thickness has a strong influence on $\Tc$, an interesting question is how the detectors' $\Tc$ compares between a thicker, higher irradiated \sspd with a thinner, lower irradiated detector that both show a similar SDE.
As elaborated in \cref{app:Tc-vs-SDE}, our data suggests that with the \qty{10}{\nm} detectors one can reach a similar SDE as with the \qty{8}{\nm} thick \sspds, while retaining a $\Tc$ of \qty{8}{\K} instead of \qty{7.5}{\K}.
This is especially useful for applications with limited cooling powers.

\begin{figure}
 \centering
 \includegraphics{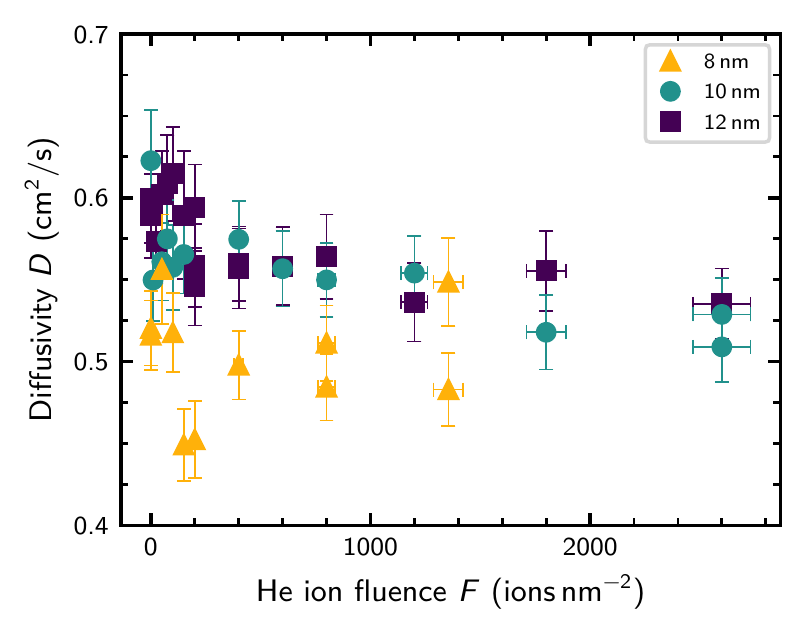}
 \caption{Quasiparticle diffusivity vs. He ion fluence, including statistical errors. 
 $D$ is almost constant within the error bars, and averaging over all fluences reveals the thickness dependence of $D$.
 One might see a slight decrease of $D$ with increasing fluence, which could be explained by the thickness reduction due to sputtering during He ion irradiation and the thickness dependence of $D$.
 }
 \label{fig:diffusivity_vs_dose}
\end{figure}

\begin{figure}
 \centering
 \includegraphics{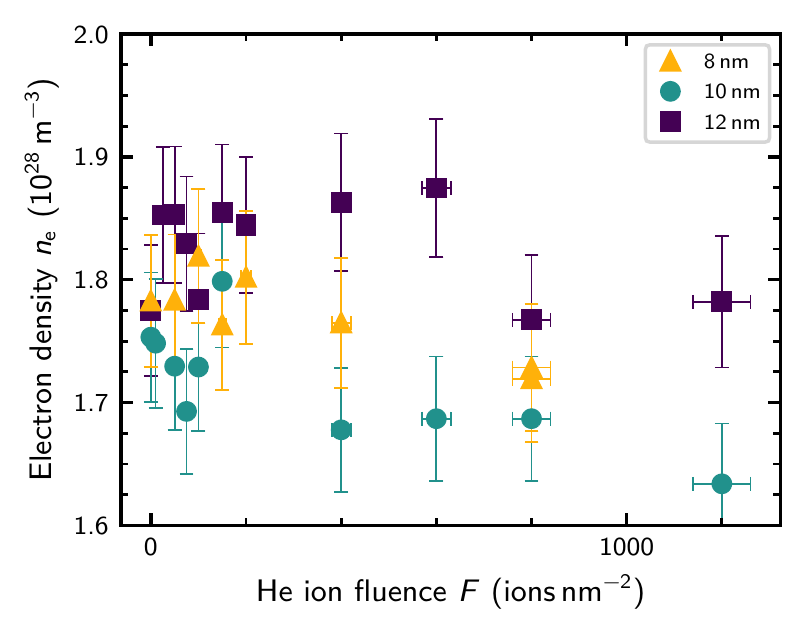}
 \caption{Electron density vs. He ion fluence, including statistical errors. Despite fluctuations between measurements, $\electrondensity$ seems almost constant. One might see a slight decrease with increasing He ion fluence, which could be explained by the thickness dependence of $\electrondensity$ and the reduction of the effective film thickness during irradiation due to sputtering.
 }
 \label{fig:electron_density_vs_dose}
\end{figure}

Next, we discuss measurements of the quasiparticle diffusivity $D$.
For this, we measured the temperature dependence of the upper critical magnetic field $B_\mathrm{c2}(T)$ by performing magneto-transport measurements while varying the temperature.
From linear fits of $B_\mathrm{c2}(T)$ close to $\Tc$, we extract the slope $\dif B_\mathrm{c2}/\dif T$ and calculate the diffusivity \cite{Semenov2009}
\begin{equation}
D 
= 
\frac{4 k_\mathrm{B}}{\pi e} 
\left[
\frac{\dif B_\mathrm{c2}}{\dif T}
\right]^{-1}_{T \rightarrow \Tc} \;.
\end{equation}
Magnetic field sweeps were also performed with the film in the normal conducting state and at constant temperature,
while measuring the Hall voltage $V_\mathrm{H}$.
Since $V_\mathrm{H}$ varies linearly with the applied magnetic field $B$ and measurement current $I$, we determine the Hall coefficient $R_\mathrm{H} = V_\mathrm{H} d_0/(I B)$ using the slope of a linear fit of the $V_\mathrm{H}(B)$ data.
From this, we estimate the electron density $n_\mathrm{e}$ according to $R_\mathrm{H} = -1/(n_\mathrm{e}e)$ within the free electron model (see also \cite{Chockalingam2008,Semenov2009, Sidorova2021}).
\Cref{fig:diffusivity_vs_dose,fig:electron_density_vs_dose} show the quasiparticle diffusivity and the electron density as a function of the He ion fluence.
Both are almost constant within the experimental error bars, although one might see a slight decrease of the diffusivity and the electron density with increasing He ion fluence.
Since we usually observe decreasing electron density and diffusivity with decreasing film thickness 
like \textcite{Sidorova2021}, this may be related to
the observed effective thickness reduction of \qty{0.94}{\nm} per \qty{1000}{\ipsn} due to sputtering during He ion irradiation.

\section{Conclusion and outlook}
In summary, we used a He ion microscope to locally tune the performance metrics of individual \sspds fabricated on the same chip.
At the same time, our results demonstrated the possibilities of using thick (up to \qty{12}{\nm}) \nbtin films and He ion irradiation to enhance performance metrics such as system detection efficiency, switching current, decay time, and operating temperature compared to \sspds of smaller thicknesses.
Thicker detectors exhibit higher optical absorption efficiency and shorter decay times as compared to similar \sspds fabricated from thinner films.
However, due to the reduction of single-photon sensitivity with detector thickness, such \sspds typically offer only small detection efficiencies.
Here, we have shown how He ion irradiation can boost the initially negligible SDE ($<\qty{0.05}{\percent}$) of \qty{12}{\nm} thick \sspds at \qty{4.5}{\K} by three orders of magnitude to \qty{55.3}{\percent}, resulting in an internal detection efficiency just within the saturated regime.
This enables the use of thicker films and the associated advantages---at temperatures reachable with standard pulse-tube or Gifford-McMahon cryocoolers.\cite{Radebaugh2009}
Furthermore, we found that by combining He ion irradiation
and detectors fabricated from thicker films, one can enhance SDE and $\Isw$ while reducing the decay time compared to non-irradiated smaller-thickness \sspds.
While reduced decay times result in increased maximum count rates, higher $\Isw$ and the associated higher detection voltage pulse imply a higher signal-to-noise ratio, which reduces the electrical noise induced timing jitter \cite{Korzh2020} and the necessary amplification of the electrical readout circuit.

Using a He ion microscope to irradiate individual detectors and cloverleaf structures on the same chip with different fluences allowed us to precisely study \sspd and film properties over He ion fluences ranging from \qty{0}{\ipsn} to \qty{2600}{\ipsn}, avoiding any errors that could arise from the high sensitivity of device properties on the exact sputtering or the subsequent fabrication process.
We found that the increase of sheet resistance with the He ion fluence can be well described by a simple physical model that includes defect generation in the \nbtin film and an effective reduction of thickness due to sputtering during He ion bombardment.
Moreover, the decrease of critical temperature with the He ion fluence can be described by combining our physical model for $\Rsheet$ with the universal scaling law from \textcite{Ivry2014}, which relates critical temperature, film thickness, and sheet resistance.
At the same time, the quasiparticle diffusivity and electron density stay almost constant for the He ion fluences studied in this work.
These magneto-transport measurements also show that irradiation of \sspds with He ions continuously changes their properties---although one can employ irradiation to enhance the \sspd performance, excessive He ion irradiation ultimately leads to a vanishing, non-detectable signal when a photon is absorbed, rendering the \sspd inoperative.
These findings could be particularly interesting for applications where \sspds are exposed to radiation and high-energy particles.\cite{Polakovic2020}

Besides the general enhancement of performance metrics of \nbtin \sspds by using thicker films combined with He ion irradiation, one can use targeted irradiation of individual devices with a He ion microscope for example in large \sspd arrays to mitigate inhomogeneities of detector performance between pixels (or even dark pixels).
This would be challenging without a post-processing technique such as site-selective He ion irradiation.
Furthermore, targeted He ion irradiation enables the optimization of detectors for different performance metrics on the same chip, also after fabrication.

\begin{acknowledgments}
The authors thank Kirill Fedorov and Stefan Appel for helpful discussions.
We gratefully acknowledge support from the German Federal Ministry of Education and Research (BMBF) via the projects PhotonQ (13N15760), SPINNING (13N16214), MARQUAND (BN105022), and ``Photonics Research Germany'' (13N14846), via the funding program ``Quantum technologies -- from basic research to market'' (16K1SQ033, 13N15855, and 13N15982), 
as well as from the German Research Foundation (DFG) under Germany's Excellence Strategy EXC-2111 (390814868) and projects INST 95/1220-1 (MQCL) and INST 95/1654-1 FUGG. 
This research is part of the ``Munich Quantum Valley'', which is supported by the Bavarian state government with funds from the ``Hightech Agenda Bayern Plus''.

\end{acknowledgments}

\appendix

\section{Simulation of optical absorption in \sspds}
\label{app:absorption-simulation}
The optical absorption in a detector provides an upper limit for its SDE.
To determine the absorption for the detectors fabricated in this work, we performed finite-difference time domain (FDTD) simulations (Ansys Lumerical).
Input parameters for these simulations are the width and thickness of the nanowire and the optical constants of the superconducting film that provides the basis for the detectors.
We controlled the thickness of the films by measuring the sputter deposition rate and selecting the deposition time accordingly.
The optical constants were measured with a variable angle spectroscopic ellipsometer (M-2000, J.A. Woollam Co.).
After detector fabrication, we evaluated the width of 22 representative detectors (Genesys ProSEM) and determined their mean wire width as listed in \cref{tab:absorption}.
Moreover, for the simulations we chose a plane-wave source with its polarization parallel to the nanowire, in line with the experiment.
\Cref{tab:absorption} shows the simulation input parameters and results for the optical absorption of the detectors of this work.
The design of all detectors consists of \qty{100}{\nm} wide wires and a fill factor of \qty{50}{\percent} as described in \cref{sec:experimental}.
The measured width of the fabricated detectors deviates from this nominal value due to slight under-/overexposure during electron-beam lithography.
This, however, does not change the reasoning within this work: 
For the same pitch, increased wire width (increased fill factor) or increased thickness both increase optical absorption and switching current, while reducing the detector's sensitivity to single photons.

\begin{table}
	\centering
	\caption{Simulation parameters and results for the absorption in the \qty{8}{\nm}, \qty{10}{\nm}, and \qty{12}{\nm} thick detectors of this work.
 $d_0$ is the nominal thickness of the detector, while $w$ represents its mean wire width. 
 $n$ and $k$ are refractive index and extinction coefficient, respectively.
 The absorption fraction $\alpha$ denotes the percentage of light that is absorbed in the detector, obtained from FDTD simulations.
}
\label{tab:absorption}
\begin{tabular}{
S
!{\qquad}
S[table-format=3.1(2), table-align-exponent = false, table-align-uncertainty = false]
!{\qquad}
S[table-format=1.2]
!{\qquad}
S[table-format=1.2]
!{\qquad}
S[table-format=2.1]
}
    \toprule
    {$d_0$ (\unit{\nm})} & 
    {$w$ (\unit{\nm})} &
    {$n$ (\unit{1})} &
    {$k$ (\unit{1})} &
    {$\alpha$ (\unit{\percent)}} \\
    \midrule
    8   & 92.6(4.0)  & 2.47  & 3.18  & 44.2 \\
    10  & 107.3(1.9) & 2.48  & 3.33  & 53.1 \\
    12  & 115.3(3.4) & 2.48  & 3.54  & 57.7 \\
    \bottomrule
\end{tabular}
\end{table}

\section{Comparison of \texorpdfstring{$\Tc$}{Tc} and SDE}
\label{app:Tc-vs-SDE}
\begin{figure}
 \centering
 \includegraphics{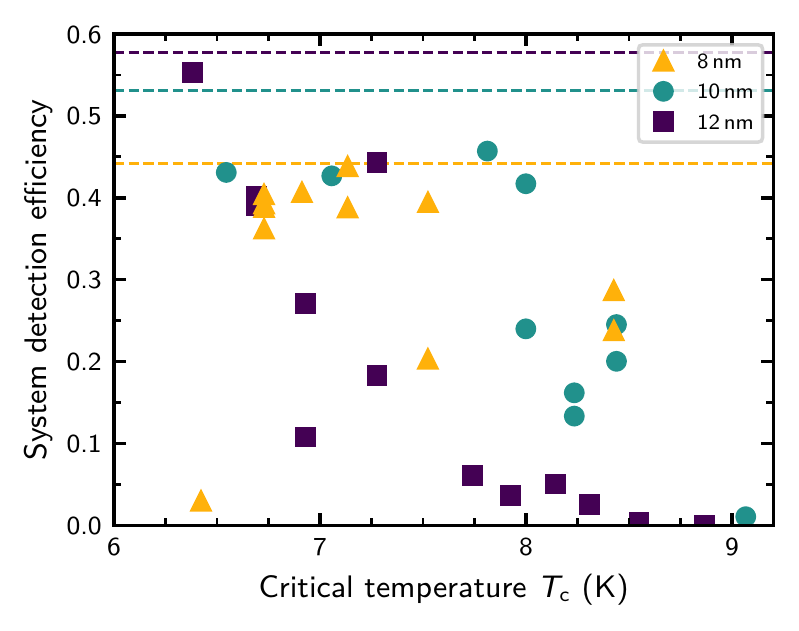}
 \caption{SDE of differently irradiated \sspds vs. $\Tc$ of corresponding CLs of three different thicknesses.
 The relative uncertainty of the SDE is \qty{2}{\percent}, the uncertainty of the temperature amounts to \qty{50}{\milli\K} (error bars not shown for clarity).
 }
 \label{fig:DE_vs_Tc_CL}
\end{figure}

The critical temperature $\Tc$ of \sspds is especially important for applications with limited cooling capabilities.
In this section, we address the question how $\Tc$ compares between thicker, higher irradiated detectors and thinner, lower irradiated \sspds that both show a similar SDE.
For this, we compare the detector's SDE with the thin-film $\Tc$ in \cref{fig:DE_vs_Tc_CL}.
The data suggests that with \qty{44}{\percent} the \qty{10}{\nm} detectors can actually reach an SDE comparable to the \qty{8}{\nm} \sspds, while retaining a $\Tc$ of \qty{8}{\K} instead of \qty{7.5}{\K}.
This can be particularly useful for applications because a higher $\Tc$ reduces the requirements for the cooling system to operate the \sspds.

\section{Fitting of \texorpdfstring{$\Tc$}{Tc} with universal scaling law}
\label{app:Tc-fitting}
\begin{figure}
 \centering
 \includegraphics{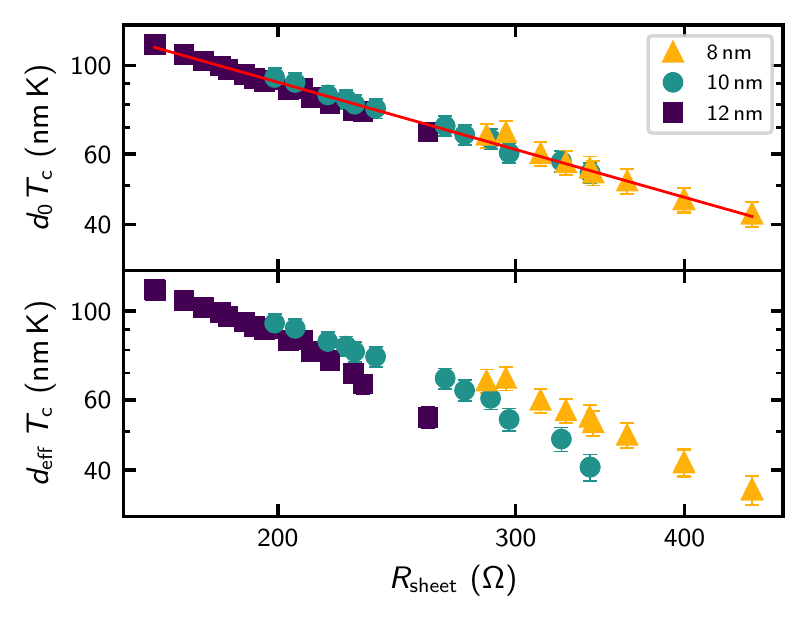}
 \caption{Critical temperature multiplied with film thickness vs. sheet resistance, including statistical errors. 
 In the upper plot the thickness of the non-irradiated film is used, while the effective thickness used in the lower plot accounts for surface sputtering and intermixing at the film/substrate interface due to He ion irradiation.
 The red line in the upper plot is a fit according to the universal scaling law introduced by \textcite{Ivry2014}, $d_0 \Tc = A \Rsheet^{-B}$.
 }
 \label{fig:d_Tc_vs_R_sheet_w_and_wo_thicknes_correction}
\end{figure}

To describe the data for the critical temperature in \cref{fig:Tc_vs_dose}, we use the universal scaling law, introduced by \textcite{Ivry2014}, 
\begin{equation}
    d \, \Tc = A \Rsheet^{-B} \;,
\end{equation}
which relates film thickness, critical temperature, and sheet resistance.
\Cref{fig:d_Tc_vs_R_sheet_w_and_wo_thicknes_correction} shows the critical temperature, multiplied with the thickness of the non-irradiated film, $d_0 \Tc$.
Evidently, this quantity exhibits a linear dependence on the sheet resistance on a log-log scale.
As the data of the differently irradiated \qty{8}{\nm},  \qty{10}{\nm}, and  \qty{12}{\nm} thick films approximately collapse on a single line, we choose one joint fitting function to determine the constants $A$ and $B$ of the universal scaling law and obtain the unitless constants $A=\num{1.44e4}$ and $B=\num{0.957}$, provided that the data for $d$, $\Tc$, and $\Rsheet$ are given in \unit{\nm}, \unit{\K}, and \unit{\ohm}, respectively.
With these parameters, we obtain the fitting functions shown in \cref{fig:Tc_vs_dose}.

It is interesting to note that the linearity of the three data sets shown in the upper part of \cref{fig:d_Tc_vs_R_sheet_w_and_wo_thicknes_correction} is lost when multiplying $\Tc$ with the effective thickness $d_\mathrm{eff} = d_0 - r_\mathrm{s}\, F$ instead of the thickness before irradiation, as shown in the lower part of \cref{fig:d_Tc_vs_R_sheet_w_and_wo_thicknes_correction}.
As introduced in \cref{sec:scaling-thin-film-with-fluence}, this reduction of the effective thickness by \qty{0.94}{\nm} per \qty{1000}{\ipsn} accounts for surface sputtering and intermixing at the film/substrate interface.
At present, we can only give a qualitative explanation why the effective thickness is important to describe the continuous increase of $\Rsheet$ in \cref{fig:R_sheet_vs_dose} and why it is not relevant for describing $\Tc$:
Via AFM measurements, we observed a surface roughening by He ion irradiation due to surface sputtering and redeposition.
Considering now a thin slab of the rough surface, parallel to the sample plane, it consists of many connected islands of \nbtin (or an oxide thereof).
On the one hand, this slab has a higher resistivity in the normal conducting state due to the voids; 
on the other hand, it should have a $\Tc$ similar to a slab without voids as long as the voids are smaller than the coherence length of the superconductor.
Of course, further investigation is necessary to better understand the role of surface sputtering and intermixing at the film/substrate interface as well as their influence on thickness, sheet resistance, and critical temperature of the thin film.

\bibliography{HIM_paper_I_main}%

\begin{thebibliography}{73}%
\makeatletter
\providecommand \@ifxundefined [1]{%
 \@ifx{#1\undefined}
}%
\providecommand \@ifnum [1]{%
 \ifnum #1\expandafter \@firstoftwo
 \else \expandafter \@secondoftwo
 \fi
}%
\providecommand \@ifx [1]{%
 \ifx #1\expandafter \@firstoftwo
 \else \expandafter \@secondoftwo
 \fi
}%
\providecommand \natexlab [1]{#1}%
\providecommand \enquote  [1]{``#1''}%
\providecommand \bibnamefont  [1]{#1}%
\providecommand \bibfnamefont [1]{#1}%
\providecommand \citenamefont [1]{#1}%
\providecommand \href@noop [0]{\@secondoftwo}%
\providecommand \href [0]{\begingroup \@sanitize@url \@href}%
\providecommand \@href[1]{\@@startlink{#1}\@@href}%
\providecommand \@@href[1]{\endgroup#1\@@endlink}%
\providecommand \@sanitize@url [0]{\catcode `\\12\catcode `\$12\catcode
  `\&12\catcode `\#12\catcode `\^12\catcode `\_12\catcode `\%12\relax}%
\providecommand \@@startlink[1]{}%
\providecommand \@@endlink[0]{}%
\providecommand \url  [0]{\begingroup\@sanitize@url \@url }%
\providecommand \@url [1]{\endgroup\@href {#1}{\urlprefix }}%
\providecommand \urlprefix  [0]{URL }%
\providecommand \Eprint [0]{\href }%
\providecommand \doibase [0]{https://doi.org/}%
\providecommand \selectlanguage [0]{\@gobble}%
\providecommand \bibinfo  [0]{\@secondoftwo}%
\providecommand \bibfield  [0]{\@secondoftwo}%
\providecommand \translation [1]{[#1]}%
\providecommand \BibitemOpen [0]{}%
\providecommand \bibitemStop [0]{}%
\providecommand \bibitemNoStop [0]{.\EOS\space}%
\providecommand \EOS [0]{\spacefactor3000\relax}%
\providecommand \BibitemShut  [1]{\csname bibitem#1\endcsname}%
\let\auto@bib@innerbib\@empty
\bibitem [{\citenamefont {Gol'tsman}\ \emph {et~al.}(2001)\citenamefont
  {Gol'tsman}, \citenamefont {Okunev}, \citenamefont {Chulkova}, \citenamefont
  {Lipatov}, \citenamefont {Semenov}, \citenamefont {Smirnov}, \citenamefont
  {Voronov}, \citenamefont {Dzardanov}, \citenamefont {Williams},\ and\
  \citenamefont {Sobolewski}}]{Goltsman2001}%
  \BibitemOpen
  \bibfield  {author} {\bibinfo {author} {\bibfnamefont {G.~N.}\ \bibnamefont
  {Gol'tsman}}, \bibinfo {author} {\bibfnamefont {O.}~\bibnamefont {Okunev}},
  \bibinfo {author} {\bibfnamefont {G.}~\bibnamefont {Chulkova}}, \bibinfo
  {author} {\bibfnamefont {A.}~\bibnamefont {Lipatov}}, \bibinfo {author}
  {\bibfnamefont {A.}~\bibnamefont {Semenov}}, \bibinfo {author} {\bibfnamefont
  {K.}~\bibnamefont {Smirnov}}, \bibinfo {author} {\bibfnamefont
  {B.}~\bibnamefont {Voronov}}, \bibinfo {author} {\bibfnamefont
  {A.}~\bibnamefont {Dzardanov}}, \bibinfo {author} {\bibfnamefont
  {C.}~\bibnamefont {Williams}},\ and\ \bibinfo {author} {\bibfnamefont
  {R.}~\bibnamefont {Sobolewski}},\ }\bibfield  {title} {\bibinfo {title}
  {Picosecond superconducting single-photon optical detector},\ }\href
  {https://doi.org/10.1063/1.1388868} {\bibfield  {journal} {\bibinfo
  {journal} {Applied Physics Letters}\ }\textbf {\bibinfo {volume} {79}},\
  \bibinfo {pages} {705} (\bibinfo {year} {2001})}\BibitemShut {NoStop}%
\bibitem [{\citenamefont {Takesue}\ \emph {et~al.}(2007)\citenamefont
  {Takesue}, \citenamefont {Nam}, \citenamefont {Zhang}, \citenamefont
  {Hadfield}, \citenamefont {Honjo}, \citenamefont {Tamaki},\ and\
  \citenamefont {Yamamoto}}]{Takesue2007}%
  \BibitemOpen
  \bibfield  {author} {\bibinfo {author} {\bibfnamefont {H.}~\bibnamefont
  {Takesue}}, \bibinfo {author} {\bibfnamefont {S.~W.}\ \bibnamefont {Nam}},
  \bibinfo {author} {\bibfnamefont {Q.}~\bibnamefont {Zhang}}, \bibinfo
  {author} {\bibfnamefont {R.~H.}\ \bibnamefont {Hadfield}}, \bibinfo {author}
  {\bibfnamefont {T.}~\bibnamefont {Honjo}}, \bibinfo {author} {\bibfnamefont
  {K.}~\bibnamefont {Tamaki}},\ and\ \bibinfo {author} {\bibfnamefont
  {Y.}~\bibnamefont {Yamamoto}},\ }\bibfield  {title} {\bibinfo {title}
  {Quantum key distribution over a 40-{{dB}} channel loss using superconducting
  single-photon detectors},\ }\href {https://doi.org/10.1038/nphoton.2007.75}
  {\bibfield  {journal} {\bibinfo  {journal} {Nature Photonics}\ }\textbf
  {\bibinfo {volume} {1}},\ \bibinfo {pages} {343} (\bibinfo {year}
  {2007})}\BibitemShut {NoStop}%
\bibitem [{\citenamefont {Chen}\ \emph {et~al.}(2022)\citenamefont {Chen},
  \citenamefont {Zhang}, \citenamefont {Liu}, \citenamefont {Jiang},
  \citenamefont {Zhao}, \citenamefont {Zhang}, \citenamefont {Chen},
  \citenamefont {Li}, \citenamefont {You}, \citenamefont {Wang}, \citenamefont
  {Chen}, \citenamefont {Wang}, \citenamefont {Zhang},\ and\ \citenamefont
  {Pan}}]{Chen2022}%
  \BibitemOpen
  \bibfield  {author} {\bibinfo {author} {\bibfnamefont {J.-P.}\ \bibnamefont
  {Chen}}, \bibinfo {author} {\bibfnamefont {C.}~\bibnamefont {Zhang}},
  \bibinfo {author} {\bibfnamefont {Y.}~\bibnamefont {Liu}}, \bibinfo {author}
  {\bibfnamefont {C.}~\bibnamefont {Jiang}}, \bibinfo {author} {\bibfnamefont
  {D.-F.}\ \bibnamefont {Zhao}}, \bibinfo {author} {\bibfnamefont {W.-J.}\
  \bibnamefont {Zhang}}, \bibinfo {author} {\bibfnamefont {F.-X.}\ \bibnamefont
  {Chen}}, \bibinfo {author} {\bibfnamefont {H.}~\bibnamefont {Li}}, \bibinfo
  {author} {\bibfnamefont {L.-X.}\ \bibnamefont {You}}, \bibinfo {author}
  {\bibfnamefont {Z.}~\bibnamefont {Wang}}, \bibinfo {author} {\bibfnamefont
  {Y.}~\bibnamefont {Chen}}, \bibinfo {author} {\bibfnamefont {X.-B.}\
  \bibnamefont {Wang}}, \bibinfo {author} {\bibfnamefont {Q.}~\bibnamefont
  {Zhang}},\ and\ \bibinfo {author} {\bibfnamefont {J.-W.}\ \bibnamefont
  {Pan}},\ }\bibfield  {title} {\bibinfo {title} {Quantum {{Key Distribution}}
  over 658 km {{Fiber}} with {{Distributed Vibration Sensing}}},\ }\href
  {https://doi.org/10.1103/PhysRevLett.128.180502} {\bibfield  {journal}
  {\bibinfo  {journal} {Physical Review Letters}\ }\textbf {\bibinfo {volume}
  {128}},\ \bibinfo {pages} {180502} (\bibinfo {year} {2022})}\BibitemShut
  {NoStop}%
\bibitem [{\citenamefont {Liu}\ \emph {et~al.}(2023)\citenamefont {Liu},
  \citenamefont {Zhang}, \citenamefont {Jiang}, \citenamefont {Chen},
  \citenamefont {Zhang}, \citenamefont {Pan}, \citenamefont {Ma}, \citenamefont
  {Dong}, \citenamefont {Xiong}, \citenamefont {Zhang}, \citenamefont {Li},
  \citenamefont {Wang}, \citenamefont {Wu}, \citenamefont {Chen}, \citenamefont
  {You}, \citenamefont {Wang}, \citenamefont {Zhang},\ and\ \citenamefont
  {Pan}}]{Liu2023}%
  \BibitemOpen
  \bibfield  {author} {\bibinfo {author} {\bibfnamefont {Y.}~\bibnamefont
  {Liu}}, \bibinfo {author} {\bibfnamefont {W.-J.}\ \bibnamefont {Zhang}},
  \bibinfo {author} {\bibfnamefont {C.}~\bibnamefont {Jiang}}, \bibinfo
  {author} {\bibfnamefont {J.-P.}\ \bibnamefont {Chen}}, \bibinfo {author}
  {\bibfnamefont {C.}~\bibnamefont {Zhang}}, \bibinfo {author} {\bibfnamefont
  {W.-X.}\ \bibnamefont {Pan}}, \bibinfo {author} {\bibfnamefont
  {D.}~\bibnamefont {Ma}}, \bibinfo {author} {\bibfnamefont {H.}~\bibnamefont
  {Dong}}, \bibinfo {author} {\bibfnamefont {J.-M.}\ \bibnamefont {Xiong}},
  \bibinfo {author} {\bibfnamefont {C.-J.}\ \bibnamefont {Zhang}}, \bibinfo
  {author} {\bibfnamefont {H.}~\bibnamefont {Li}}, \bibinfo {author}
  {\bibfnamefont {R.-C.}\ \bibnamefont {Wang}}, \bibinfo {author}
  {\bibfnamefont {J.}~\bibnamefont {Wu}}, \bibinfo {author} {\bibfnamefont
  {T.-Y.}\ \bibnamefont {Chen}}, \bibinfo {author} {\bibfnamefont
  {L.}~\bibnamefont {You}}, \bibinfo {author} {\bibfnamefont {X.-B.}\
  \bibnamefont {Wang}}, \bibinfo {author} {\bibfnamefont {Q.}~\bibnamefont
  {Zhang}},\ and\ \bibinfo {author} {\bibfnamefont {J.-W.}\ \bibnamefont
  {Pan}},\ }\href@noop {} {\bibinfo {title} {Experimental {{Twin-Field Quantum
  Key Distribution Over}} 1000 km {{Fiber Distance}}}} (\bibinfo {year}
  {2023}),\ \Eprint {https://arxiv.org/abs/2303.15795} {arxiv:2303.15795
  [quant-ph]} \BibitemShut {NoStop}%
\bibitem [{\citenamefont {Bussi{\`e}res}\ \emph {et~al.}(2014)\citenamefont
  {Bussi{\`e}res}, \citenamefont {Clausen}, \citenamefont {Tiranov},
  \citenamefont {Korzh}, \citenamefont {Verma}, \citenamefont {Nam},
  \citenamefont {Marsili}, \citenamefont {Ferrier}, \citenamefont {Goldner},
  \citenamefont {Herrmann}, \citenamefont {Silberhorn}, \citenamefont {Sohler},
  \citenamefont {Afzelius},\ and\ \citenamefont {Gisin}}]{Bussieres2014}%
  \BibitemOpen
  \bibfield  {author} {\bibinfo {author} {\bibfnamefont {F.}~\bibnamefont
  {Bussi{\`e}res}}, \bibinfo {author} {\bibfnamefont {C.}~\bibnamefont
  {Clausen}}, \bibinfo {author} {\bibfnamefont {A.}~\bibnamefont {Tiranov}},
  \bibinfo {author} {\bibfnamefont {B.}~\bibnamefont {Korzh}}, \bibinfo
  {author} {\bibfnamefont {V.~B.}\ \bibnamefont {Verma}}, \bibinfo {author}
  {\bibfnamefont {S.~W.}\ \bibnamefont {Nam}}, \bibinfo {author} {\bibfnamefont
  {F.}~\bibnamefont {Marsili}}, \bibinfo {author} {\bibfnamefont
  {A.}~\bibnamefont {Ferrier}}, \bibinfo {author} {\bibfnamefont
  {P.}~\bibnamefont {Goldner}}, \bibinfo {author} {\bibfnamefont
  {H.}~\bibnamefont {Herrmann}}, \bibinfo {author} {\bibfnamefont
  {C.}~\bibnamefont {Silberhorn}}, \bibinfo {author} {\bibfnamefont
  {W.}~\bibnamefont {Sohler}}, \bibinfo {author} {\bibfnamefont
  {M.}~\bibnamefont {Afzelius}},\ and\ \bibinfo {author} {\bibfnamefont
  {N.}~\bibnamefont {Gisin}},\ }\bibfield  {title} {\bibinfo {title} {Quantum
  teleportation from a telecom-wavelength photon to a solid-state quantum
  memory},\ }\href {https://doi.org/10.1038/nphoton.2014.215} {\bibfield
  {journal} {\bibinfo  {journal} {Nature Photonics}\ }\textbf {\bibinfo
  {volume} {8}},\ \bibinfo {pages} {775} (\bibinfo {year} {2014})}\BibitemShut
  {NoStop}%
\bibitem [{\citenamefont {McCarthy}\ \emph {et~al.}(2013)\citenamefont
  {McCarthy}, \citenamefont {Krichel}, \citenamefont {Gemmell}, \citenamefont
  {Ren}, \citenamefont {Tanner}, \citenamefont {Dorenbos}, \citenamefont
  {Zwiller}, \citenamefont {Hadfield},\ and\ \citenamefont
  {Buller}}]{McCarthy2013}%
  \BibitemOpen
  \bibfield  {author} {\bibinfo {author} {\bibfnamefont {A.}~\bibnamefont
  {McCarthy}}, \bibinfo {author} {\bibfnamefont {N.~J.}\ \bibnamefont
  {Krichel}}, \bibinfo {author} {\bibfnamefont {N.~R.}\ \bibnamefont
  {Gemmell}}, \bibinfo {author} {\bibfnamefont {X.}~\bibnamefont {Ren}},
  \bibinfo {author} {\bibfnamefont {M.~G.}\ \bibnamefont {Tanner}}, \bibinfo
  {author} {\bibfnamefont {S.~N.}\ \bibnamefont {Dorenbos}}, \bibinfo {author}
  {\bibfnamefont {V.}~\bibnamefont {Zwiller}}, \bibinfo {author} {\bibfnamefont
  {R.~H.}\ \bibnamefont {Hadfield}},\ and\ \bibinfo {author} {\bibfnamefont
  {G.~S.}\ \bibnamefont {Buller}},\ }\bibfield  {title} {\bibinfo {title}
  {Kilometer-range, high resolution depth imaging via 1560 nm wavelength
  single-photon detection},\ }\href {https://doi.org/10.1364/OE.21.008904}
  {\bibfield  {journal} {\bibinfo  {journal} {Optics Express}\ }\textbf
  {\bibinfo {volume} {21}},\ \bibinfo {pages} {8904} (\bibinfo {year}
  {2013})}\BibitemShut {NoStop}%
\bibitem [{\citenamefont {Grein}\ \emph {et~al.}(2015)\citenamefont {Grein},
  \citenamefont {Kerman}, \citenamefont {Dauler}, \citenamefont {Willis},
  \citenamefont {Romkey}, \citenamefont {Molnar}, \citenamefont {Robinson},
  \citenamefont {Murphy},\ and\ \citenamefont {Boroson}}]{Grein2015}%
  \BibitemOpen
  \bibfield  {author} {\bibinfo {author} {\bibfnamefont {M.~E.}\ \bibnamefont
  {Grein}}, \bibinfo {author} {\bibfnamefont {A.~J.}\ \bibnamefont {Kerman}},
  \bibinfo {author} {\bibfnamefont {E.~A.}\ \bibnamefont {Dauler}}, \bibinfo
  {author} {\bibfnamefont {M.~M.}\ \bibnamefont {Willis}}, \bibinfo {author}
  {\bibfnamefont {B.}~\bibnamefont {Romkey}}, \bibinfo {author} {\bibfnamefont
  {R.~J.}\ \bibnamefont {Molnar}}, \bibinfo {author} {\bibfnamefont {B.~S.}\
  \bibnamefont {Robinson}}, \bibinfo {author} {\bibfnamefont {D.~V.}\
  \bibnamefont {Murphy}},\ and\ \bibinfo {author} {\bibfnamefont {D.~M.}\
  \bibnamefont {Boroson}},\ }\bibfield  {title} {\bibinfo {title} {An optical
  receiver for the {{Lunar Laser Communication Demonstration}} based on
  photon-counting superconducting nanowires},\ }\href
  {https://doi.org/10.1117/12.2179781} {\bibfield  {journal} {\bibinfo
  {journal} {Advanced Photon Counting Techniques IX}\ }\textbf {\bibinfo
  {volume} {9492}},\ \bibinfo {pages} {949208} (\bibinfo {year}
  {2015})}\BibitemShut {NoStop}%
\bibitem [{\citenamefont {Takesue}\ \emph {et~al.}(2015)\citenamefont
  {Takesue}, \citenamefont {Dyer}, \citenamefont {Stevens}, \citenamefont
  {Verma}, \citenamefont {Mirin},\ and\ \citenamefont {Nam}}]{Takesue2015}%
  \BibitemOpen
  \bibfield  {author} {\bibinfo {author} {\bibfnamefont {H.}~\bibnamefont
  {Takesue}}, \bibinfo {author} {\bibfnamefont {S.~D.}\ \bibnamefont {Dyer}},
  \bibinfo {author} {\bibfnamefont {M.~J.}\ \bibnamefont {Stevens}}, \bibinfo
  {author} {\bibfnamefont {V.}~\bibnamefont {Verma}}, \bibinfo {author}
  {\bibfnamefont {R.~P.}\ \bibnamefont {Mirin}},\ and\ \bibinfo {author}
  {\bibfnamefont {S.~W.}\ \bibnamefont {Nam}},\ }\bibfield  {title} {\bibinfo
  {title} {Quantum teleportation over 100 km of fiber using highly efficient
  superconducting nanowire single-photon detectors},\ }\href
  {https://doi.org/10.1364/OPTICA.2.000832} {\bibfield  {journal} {\bibinfo
  {journal} {Optica}\ }\textbf {\bibinfo {volume} {2}},\ \bibinfo {pages} {832}
  (\bibinfo {year} {2015})}\BibitemShut {NoStop}%
\bibitem [{\citenamefont {Valivarthi}\ \emph {et~al.}(2016)\citenamefont
  {Valivarthi}, \citenamefont {Puigibert}, \citenamefont {Zhou}, \citenamefont
  {Aguilar}, \citenamefont {Verma}, \citenamefont {Marsili}, \citenamefont
  {Shaw}, \citenamefont {Nam}, \citenamefont {Oblak},\ and\ \citenamefont
  {Tittel}}]{Valivarthi2016}%
  \BibitemOpen
  \bibfield  {author} {\bibinfo {author} {\bibfnamefont {R.}~\bibnamefont
  {Valivarthi}}, \bibinfo {author} {\bibfnamefont {M.~G.}\ \bibnamefont
  {Puigibert}}, \bibinfo {author} {\bibfnamefont {Q.}~\bibnamefont {Zhou}},
  \bibinfo {author} {\bibfnamefont {G.~H.}\ \bibnamefont {Aguilar}}, \bibinfo
  {author} {\bibfnamefont {V.~B.}\ \bibnamefont {Verma}}, \bibinfo {author}
  {\bibfnamefont {F.}~\bibnamefont {Marsili}}, \bibinfo {author} {\bibfnamefont
  {M.~D.}\ \bibnamefont {Shaw}}, \bibinfo {author} {\bibfnamefont {S.~W.}\
  \bibnamefont {Nam}}, \bibinfo {author} {\bibfnamefont {D.}~\bibnamefont
  {Oblak}},\ and\ \bibinfo {author} {\bibfnamefont {W.}~\bibnamefont
  {Tittel}},\ }\bibfield  {title} {\bibinfo {title} {Quantum teleportation
  across a metropolitan fibre network},\ }\href
  {https://doi.org/10.1038/nphoton.2016.180} {\bibfield  {journal} {\bibinfo
  {journal} {Nature Photonics}\ }\textbf {\bibinfo {volume} {10}},\ \bibinfo
  {pages} {676} (\bibinfo {year} {2016})}\BibitemShut {NoStop}%
\bibitem [{\citenamefont {Zhang}\ \emph {et~al.}(2003)\citenamefont {Zhang},
  \citenamefont {Boiadjieva}, \citenamefont {Chulkova}, \citenamefont
  {Deslandes}, \citenamefont {Gol'tsman}, \citenamefont {Korneev},
  \citenamefont {Kouminov}, \citenamefont {Leibowitz}, \citenamefont {Lo},
  \citenamefont {Malinsky}, \citenamefont {Okunev}, \citenamefont {Pearlman},
  \citenamefont {Slysz}, \citenamefont {Smirnov}, \citenamefont {Tsao},
  \citenamefont {Verevkin}, \citenamefont {Voronov}, \citenamefont {Wilsher},\
  and\ \citenamefont {Sobolewski}}]{Zhang2003}%
  \BibitemOpen
  \bibfield  {author} {\bibinfo {author} {\bibfnamefont {J.}~\bibnamefont
  {Zhang}}, \bibinfo {author} {\bibfnamefont {N.}~\bibnamefont {Boiadjieva}},
  \bibinfo {author} {\bibfnamefont {G.}~\bibnamefont {Chulkova}}, \bibinfo
  {author} {\bibfnamefont {H.}~\bibnamefont {Deslandes}}, \bibinfo {author}
  {\bibfnamefont {G.}~\bibnamefont {Gol'tsman}}, \bibinfo {author}
  {\bibfnamefont {A.}~\bibnamefont {Korneev}}, \bibinfo {author} {\bibfnamefont
  {P.}~\bibnamefont {Kouminov}}, \bibinfo {author} {\bibfnamefont
  {M.}~\bibnamefont {Leibowitz}}, \bibinfo {author} {\bibfnamefont
  {W.}~\bibnamefont {Lo}}, \bibinfo {author} {\bibfnamefont {R.}~\bibnamefont
  {Malinsky}}, \bibinfo {author} {\bibfnamefont {O.}~\bibnamefont {Okunev}},
  \bibinfo {author} {\bibfnamefont {A.}~\bibnamefont {Pearlman}}, \bibinfo
  {author} {\bibfnamefont {W.}~\bibnamefont {Slysz}}, \bibinfo {author}
  {\bibfnamefont {K.}~\bibnamefont {Smirnov}}, \bibinfo {author} {\bibfnamefont
  {C.}~\bibnamefont {Tsao}}, \bibinfo {author} {\bibfnamefont {A.}~\bibnamefont
  {Verevkin}}, \bibinfo {author} {\bibfnamefont {B.}~\bibnamefont {Voronov}},
  \bibinfo {author} {\bibfnamefont {K.}~\bibnamefont {Wilsher}},\ and\ \bibinfo
  {author} {\bibfnamefont {R.}~\bibnamefont {Sobolewski}},\ }\bibfield  {title}
  {\bibinfo {title} {Noninvasive {{CMOS}} circuit testing with {{NbN}}
  superconducting single-photon detectors},\ }\href
  {https://doi.org/10.1049/el:20030710} {\bibfield  {journal} {\bibinfo
  {journal} {Electronics Letters}\ }\textbf {\bibinfo {volume} {39}},\ \bibinfo
  {pages} {1086} (\bibinfo {year} {2003})}\BibitemShut {NoStop}%
\bibitem [{\citenamefont {Tanner}\ \emph {et~al.}(2011)\citenamefont {Tanner},
  \citenamefont {Dyer}, \citenamefont {Baek}, \citenamefont {Hadfield},\ and\
  \citenamefont {Woo~Nam}}]{Tanner2011}%
  \BibitemOpen
  \bibfield  {author} {\bibinfo {author} {\bibfnamefont {M.~G.}\ \bibnamefont
  {Tanner}}, \bibinfo {author} {\bibfnamefont {S.~D.}\ \bibnamefont {Dyer}},
  \bibinfo {author} {\bibfnamefont {B.}~\bibnamefont {Baek}}, \bibinfo {author}
  {\bibfnamefont {R.~H.}\ \bibnamefont {Hadfield}},\ and\ \bibinfo {author}
  {\bibfnamefont {S.}~\bibnamefont {Woo~Nam}},\ }\bibfield  {title} {\bibinfo
  {title} {High-resolution single-mode fiber-optic distributed {{Raman}} sensor
  for absolute temperature measurement using superconducting nanowire
  single-photon detectors},\ }\href {https://doi.org/10.1063/1.3656702}
  {\bibfield  {journal} {\bibinfo  {journal} {Applied Physics Letters}\
  }\textbf {\bibinfo {volume} {99}},\ \bibinfo {pages} {201110} (\bibinfo
  {year} {2011})}\BibitemShut {NoStop}%
\bibitem [{\citenamefont {Itzler}\ \emph {et~al.}(2011)\citenamefont {Itzler},
  \citenamefont {Jiang}, \citenamefont {Entwistle}, \citenamefont {Slomkowski},
  \citenamefont {Tosi}, \citenamefont {Acerbi}, \citenamefont {Zappa},\ and\
  \citenamefont {Cova}}]{Itzler2011}%
  \BibitemOpen
  \bibfield  {author} {\bibinfo {author} {\bibfnamefont {M.~A.}\ \bibnamefont
  {Itzler}}, \bibinfo {author} {\bibfnamefont {X.}~\bibnamefont {Jiang}},
  \bibinfo {author} {\bibfnamefont {M.}~\bibnamefont {Entwistle}}, \bibinfo
  {author} {\bibfnamefont {K.}~\bibnamefont {Slomkowski}}, \bibinfo {author}
  {\bibfnamefont {A.}~\bibnamefont {Tosi}}, \bibinfo {author} {\bibfnamefont
  {F.}~\bibnamefont {Acerbi}}, \bibinfo {author} {\bibfnamefont
  {F.}~\bibnamefont {Zappa}},\ and\ \bibinfo {author} {\bibfnamefont
  {S.}~\bibnamefont {Cova}},\ }\bibfield  {title} {\bibinfo {title} {Advances
  in {{InGaAsP-based}} avalanche diode single photon detectors},\ }\href
  {https://doi.org/10.1080/09500340.2010.547262} {\bibfield  {journal}
  {\bibinfo  {journal} {Journal of Modern Optics}\ }\textbf {\bibinfo {volume}
  {58}},\ \bibinfo {pages} {174} (\bibinfo {year} {2011})}\BibitemShut
  {NoStop}%
\bibitem [{\citenamefont {Marsili}\ \emph {et~al.}(2012)\citenamefont
  {Marsili}, \citenamefont {Bellei}, \citenamefont {Najafi}, \citenamefont
  {Dane}, \citenamefont {Dauler}, \citenamefont {Molnar},\ and\ \citenamefont
  {Berggren}}]{Marsili2012a}%
  \BibitemOpen
  \bibfield  {author} {\bibinfo {author} {\bibfnamefont {F.}~\bibnamefont
  {Marsili}}, \bibinfo {author} {\bibfnamefont {F.}~\bibnamefont {Bellei}},
  \bibinfo {author} {\bibfnamefont {F.}~\bibnamefont {Najafi}}, \bibinfo
  {author} {\bibfnamefont {A.~E.}\ \bibnamefont {Dane}}, \bibinfo {author}
  {\bibfnamefont {E.~A.}\ \bibnamefont {Dauler}}, \bibinfo {author}
  {\bibfnamefont {R.~J.}\ \bibnamefont {Molnar}},\ and\ \bibinfo {author}
  {\bibfnamefont {K.~K.}\ \bibnamefont {Berggren}},\ }\bibfield  {title}
  {\bibinfo {title} {Efficient {{Single Photon Detection}} from 500nm to
  5\textmu m {{Wavelength}}},\ }\href {https://doi.org/10.1021/nl302245n}
  {\bibfield  {journal} {\bibinfo  {journal} {Nano Letters}\ }\textbf {\bibinfo
  {volume} {12}},\ \bibinfo {pages} {4799} (\bibinfo {year}
  {2012})}\BibitemShut {NoStop}%
\bibitem [{\citenamefont {Korneev}\ \emph {et~al.}(2012)\citenamefont
  {Korneev}, \citenamefont {Korneeva}, \citenamefont {Florya}, \citenamefont
  {Voronov},\ and\ \citenamefont {Goltsman}}]{Korneev2012}%
  \BibitemOpen
  \bibfield  {author} {\bibinfo {author} {\bibfnamefont {A.}~\bibnamefont
  {Korneev}}, \bibinfo {author} {\bibfnamefont {{\relax Yu}.}~\bibnamefont
  {Korneeva}}, \bibinfo {author} {\bibfnamefont {I.}~\bibnamefont {Florya}},
  \bibinfo {author} {\bibfnamefont {B.}~\bibnamefont {Voronov}},\ and\ \bibinfo
  {author} {\bibfnamefont {G.}~\bibnamefont {Goltsman}},\ }\bibfield  {title}
  {\bibinfo {title} {{{NbN Nanowire Superconducting Single-Photon Detector}}
  for {{Mid-Infrared}}},\ }\href {https://doi.org/10.1016/j.phpro.2012.06.215}
  {\bibfield  {journal} {\bibinfo  {journal} {Physics Procedia}\ }\textbf
  {\bibinfo {volume} {36}},\ \bibinfo {pages} {72} (\bibinfo {year}
  {2012})}\BibitemShut {NoStop}%
\bibitem [{\citenamefont {Shibata}\ \emph {et~al.}(2015)\citenamefont
  {Shibata}, \citenamefont {Shimizu}, \citenamefont {Takesue},\ and\
  \citenamefont {Tokura}}]{Shibata2015}%
  \BibitemOpen
  \bibfield  {author} {\bibinfo {author} {\bibfnamefont {H.}~\bibnamefont
  {Shibata}}, \bibinfo {author} {\bibfnamefont {K.}~\bibnamefont {Shimizu}},
  \bibinfo {author} {\bibfnamefont {H.}~\bibnamefont {Takesue}},\ and\ \bibinfo
  {author} {\bibfnamefont {Y.}~\bibnamefont {Tokura}},\ }\bibfield  {title}
  {\bibinfo {title} {Ultimate low system dark-count rate for superconducting
  nanowire single-photon detector},\ }\href
  {https://doi.org/10.1364/ol.40.003428} {\bibfield  {journal} {\bibinfo
  {journal} {Optics Letters}\ }\textbf {\bibinfo {volume} {40}},\ \bibinfo
  {pages} {3428} (\bibinfo {year} {2015})}\BibitemShut {NoStop}%
\bibitem [{\citenamefont {Korzh}\ \emph {et~al.}(2020)\citenamefont {Korzh},
  \citenamefont {Zhao}, \citenamefont {Allmaras}, \citenamefont {Frasca},
  \citenamefont {Autry}, \citenamefont {Bersin}, \citenamefont {Beyer},
  \citenamefont {Briggs}, \citenamefont {Bumble}, \citenamefont {Colangelo},
  \citenamefont {Crouch}, \citenamefont {Dane}, \citenamefont {Gerrits},
  \citenamefont {Lita}, \citenamefont {Marsili}, \citenamefont {Moody},
  \citenamefont {Pe{\~n}a}, \citenamefont {Ramirez}, \citenamefont {Rezac},
  \citenamefont {Sinclair}, \citenamefont {Stevens}, \citenamefont {Velasco},
  \citenamefont {Verma}, \citenamefont {Wollman}, \citenamefont {Xie},
  \citenamefont {Zhu}, \citenamefont {Hale}, \citenamefont {Spiropulu},
  \citenamefont {Silverman}, \citenamefont {Mirin}, \citenamefont {Nam},
  \citenamefont {Kozorezov}, \citenamefont {Shaw},\ and\ \citenamefont
  {Berggren}}]{Korzh2020}%
  \BibitemOpen
  \bibfield  {author} {\bibinfo {author} {\bibfnamefont {B.}~\bibnamefont
  {Korzh}}, \bibinfo {author} {\bibfnamefont {Q.~Y.}\ \bibnamefont {Zhao}},
  \bibinfo {author} {\bibfnamefont {J.~P.}\ \bibnamefont {Allmaras}}, \bibinfo
  {author} {\bibfnamefont {S.}~\bibnamefont {Frasca}}, \bibinfo {author}
  {\bibfnamefont {T.~M.}\ \bibnamefont {Autry}}, \bibinfo {author}
  {\bibfnamefont {E.~A.}\ \bibnamefont {Bersin}}, \bibinfo {author}
  {\bibfnamefont {A.~D.}\ \bibnamefont {Beyer}}, \bibinfo {author}
  {\bibfnamefont {R.~M.}\ \bibnamefont {Briggs}}, \bibinfo {author}
  {\bibfnamefont {B.}~\bibnamefont {Bumble}}, \bibinfo {author} {\bibfnamefont
  {M.}~\bibnamefont {Colangelo}}, \bibinfo {author} {\bibfnamefont {G.~M.}\
  \bibnamefont {Crouch}}, \bibinfo {author} {\bibfnamefont {A.~E.}\
  \bibnamefont {Dane}}, \bibinfo {author} {\bibfnamefont {T.}~\bibnamefont
  {Gerrits}}, \bibinfo {author} {\bibfnamefont {A.~E.}\ \bibnamefont {Lita}},
  \bibinfo {author} {\bibfnamefont {F.}~\bibnamefont {Marsili}}, \bibinfo
  {author} {\bibfnamefont {G.}~\bibnamefont {Moody}}, \bibinfo {author}
  {\bibfnamefont {C.}~\bibnamefont {Pe{\~n}a}}, \bibinfo {author}
  {\bibfnamefont {E.}~\bibnamefont {Ramirez}}, \bibinfo {author} {\bibfnamefont
  {J.~D.}\ \bibnamefont {Rezac}}, \bibinfo {author} {\bibfnamefont
  {N.}~\bibnamefont {Sinclair}}, \bibinfo {author} {\bibfnamefont {M.~J.}\
  \bibnamefont {Stevens}}, \bibinfo {author} {\bibfnamefont {A.~E.}\
  \bibnamefont {Velasco}}, \bibinfo {author} {\bibfnamefont {V.~B.}\
  \bibnamefont {Verma}}, \bibinfo {author} {\bibfnamefont {E.~E.}\ \bibnamefont
  {Wollman}}, \bibinfo {author} {\bibfnamefont {S.}~\bibnamefont {Xie}},
  \bibinfo {author} {\bibfnamefont {D.}~\bibnamefont {Zhu}}, \bibinfo {author}
  {\bibfnamefont {P.~D.}\ \bibnamefont {Hale}}, \bibinfo {author}
  {\bibfnamefont {M.}~\bibnamefont {Spiropulu}}, \bibinfo {author}
  {\bibfnamefont {K.~L.}\ \bibnamefont {Silverman}}, \bibinfo {author}
  {\bibfnamefont {R.~P.}\ \bibnamefont {Mirin}}, \bibinfo {author}
  {\bibfnamefont {S.~W.}\ \bibnamefont {Nam}}, \bibinfo {author} {\bibfnamefont
  {A.~G.}\ \bibnamefont {Kozorezov}}, \bibinfo {author} {\bibfnamefont {M.~D.}\
  \bibnamefont {Shaw}},\ and\ \bibinfo {author} {\bibfnamefont {K.~K.}\
  \bibnamefont {Berggren}},\ }\bibfield  {title} {\bibinfo {title}
  {Demonstration of sub-3 ps temporal resolution with a superconducting
  nanowire single-photon detector},\ }\href
  {https://doi.org/10.1038/s41566-020-0589-x} {\bibfield  {journal} {\bibinfo
  {journal} {Nature Photonics}\ }\textbf {\bibinfo {volume} {14}},\ \bibinfo
  {pages} {250} (\bibinfo {year} {2020})}\BibitemShut {NoStop}%
\bibitem [{\citenamefont {Natarajan}\ \emph {et~al.}(2012)\citenamefont
  {Natarajan}, \citenamefont {Tanner},\ and\ \citenamefont
  {Hadfield}}]{Natarajan2012}%
  \BibitemOpen
  \bibfield  {author} {\bibinfo {author} {\bibfnamefont {C.~M.}\ \bibnamefont
  {Natarajan}}, \bibinfo {author} {\bibfnamefont {M.~G.}\ \bibnamefont
  {Tanner}},\ and\ \bibinfo {author} {\bibfnamefont {R.~H.}\ \bibnamefont
  {Hadfield}},\ }\bibfield  {title} {\bibinfo {title} {Superconducting nanowire
  single-photon detectors: Physics and applications},\ }\href
  {https://doi.org/10.1088/0953-2048/25/6/063001} {\bibfield  {journal}
  {\bibinfo  {journal} {Superconductor Science and Technology}\ }\textbf
  {\bibinfo {volume} {25}},\ \bibinfo {pages} {063001} (\bibinfo {year}
  {2012})}\BibitemShut {NoStop}%
\bibitem [{\citenamefont {Ferrari}\ \emph {et~al.}(2018)\citenamefont
  {Ferrari}, \citenamefont {Schuck},\ and\ \citenamefont
  {Pernice}}]{Ferrari2018}%
  \BibitemOpen
  \bibfield  {author} {\bibinfo {author} {\bibfnamefont {S.}~\bibnamefont
  {Ferrari}}, \bibinfo {author} {\bibfnamefont {C.}~\bibnamefont {Schuck}},\
  and\ \bibinfo {author} {\bibfnamefont {W.}~\bibnamefont {Pernice}},\
  }\bibfield  {title} {\bibinfo {title} {Waveguide-integrated superconducting
  nanowire single-photon detectors},\ }\href
  {https://doi.org/10.1515/nanoph-2018-0059} {\bibfield  {journal} {\bibinfo
  {journal} {Nanophotonics}\ }\textbf {\bibinfo {volume} {7}},\ \bibinfo
  {pages} {1725} (\bibinfo {year} {2018})}\BibitemShut {NoStop}%
\bibitem [{\citenamefont {Sprengers}\ \emph {et~al.}(2011)\citenamefont
  {Sprengers}, \citenamefont {Gaggero}, \citenamefont {Sahin}, \citenamefont
  {Jahanmirinejad}, \citenamefont {Frucci}, \citenamefont {Mattioli},
  \citenamefont {Leoni}, \citenamefont {Beetz}, \citenamefont {Lermer},
  \citenamefont {Kamp}, \citenamefont {H{\"o}fling}, \citenamefont {Sanjines},\
  and\ \citenamefont {Fiore}}]{Sprengers2011}%
  \BibitemOpen
  \bibfield  {author} {\bibinfo {author} {\bibfnamefont {J.~P.}\ \bibnamefont
  {Sprengers}}, \bibinfo {author} {\bibfnamefont {A.}~\bibnamefont {Gaggero}},
  \bibinfo {author} {\bibfnamefont {D.}~\bibnamefont {Sahin}}, \bibinfo
  {author} {\bibfnamefont {S.}~\bibnamefont {Jahanmirinejad}}, \bibinfo
  {author} {\bibfnamefont {G.}~\bibnamefont {Frucci}}, \bibinfo {author}
  {\bibfnamefont {F.}~\bibnamefont {Mattioli}}, \bibinfo {author}
  {\bibfnamefont {R.}~\bibnamefont {Leoni}}, \bibinfo {author} {\bibfnamefont
  {J.}~\bibnamefont {Beetz}}, \bibinfo {author} {\bibfnamefont
  {M.}~\bibnamefont {Lermer}}, \bibinfo {author} {\bibfnamefont
  {M.}~\bibnamefont {Kamp}}, \bibinfo {author} {\bibfnamefont {S.}~\bibnamefont
  {H{\"o}fling}}, \bibinfo {author} {\bibfnamefont {R.}~\bibnamefont
  {Sanjines}},\ and\ \bibinfo {author} {\bibfnamefont {A.}~\bibnamefont
  {Fiore}},\ }\bibfield  {title} {\bibinfo {title} {Waveguide superconducting
  single-photon detectors for integrated quantum photonic circuits},\ }\href
  {https://doi.org/10.1063/1.3657518} {\bibfield  {journal} {\bibinfo
  {journal} {Applied Physics Letters}\ }\textbf {\bibinfo {volume} {99}},\
  \bibinfo {pages} {181110} (\bibinfo {year} {2011})}\BibitemShut {NoStop}%
\bibitem [{\citenamefont {Reithmaier}\ \emph {et~al.}(2013)\citenamefont
  {Reithmaier}, \citenamefont {Lichtmannecker}, \citenamefont {Reichert},
  \citenamefont {Hasch}, \citenamefont {M{\"u}ller}, \citenamefont {Bichler},
  \citenamefont {Gross},\ and\ \citenamefont {Finley}}]{Reithmaier2013}%
  \BibitemOpen
  \bibfield  {author} {\bibinfo {author} {\bibfnamefont {G.}~\bibnamefont
  {Reithmaier}}, \bibinfo {author} {\bibfnamefont {S.}~\bibnamefont
  {Lichtmannecker}}, \bibinfo {author} {\bibfnamefont {T.}~\bibnamefont
  {Reichert}}, \bibinfo {author} {\bibfnamefont {P.}~\bibnamefont {Hasch}},
  \bibinfo {author} {\bibfnamefont {K.}~\bibnamefont {M{\"u}ller}}, \bibinfo
  {author} {\bibfnamefont {M.}~\bibnamefont {Bichler}}, \bibinfo {author}
  {\bibfnamefont {R.}~\bibnamefont {Gross}},\ and\ \bibinfo {author}
  {\bibfnamefont {J.~J.}\ \bibnamefont {Finley}},\ }\bibfield  {title}
  {\bibinfo {title} {On-chip time resolved detection of quantum dot emission
  using integrated superconducting single photon detectors},\ }\href
  {https://doi.org/10.1038/srep01901} {\bibfield  {journal} {\bibinfo
  {journal} {Scientific Reports}\ }\textbf {\bibinfo {volume} {3}},\ \bibinfo
  {pages} {1901} (\bibinfo {year} {2013})}\BibitemShut {NoStop}%
\bibitem [{\citenamefont {Reithmaier}\ \emph {et~al.}(2015)\citenamefont
  {Reithmaier}, \citenamefont {Kaniber}, \citenamefont {Flassig}, \citenamefont
  {Lichtmannecker}, \citenamefont {M{\"u}ller}, \citenamefont {Andrejew},
  \citenamefont {Vu{\v c}kovi{\'c}}, \citenamefont {Gross},\ and\ \citenamefont
  {Finley}}]{Reithmaier2015}%
  \BibitemOpen
  \bibfield  {author} {\bibinfo {author} {\bibfnamefont {G.}~\bibnamefont
  {Reithmaier}}, \bibinfo {author} {\bibfnamefont {M.}~\bibnamefont {Kaniber}},
  \bibinfo {author} {\bibfnamefont {F.}~\bibnamefont {Flassig}}, \bibinfo
  {author} {\bibfnamefont {S.}~\bibnamefont {Lichtmannecker}}, \bibinfo
  {author} {\bibfnamefont {K.}~\bibnamefont {M{\"u}ller}}, \bibinfo {author}
  {\bibfnamefont {A.}~\bibnamefont {Andrejew}}, \bibinfo {author}
  {\bibfnamefont {J.}~\bibnamefont {Vu{\v c}kovi{\'c}}}, \bibinfo {author}
  {\bibfnamefont {R.}~\bibnamefont {Gross}},\ and\ \bibinfo {author}
  {\bibfnamefont {J.~J.}\ \bibnamefont {Finley}},\ }\bibfield  {title}
  {\bibinfo {title} {On-{{Chip Generation}}, {{Routing}}, and {{Detection}} of
  {{Resonance Fluorescence}}},\ }\href
  {https://doi.org/10.1021/acs.nanolett.5b01444} {\bibfield  {journal}
  {\bibinfo  {journal} {Nano Letters}\ }\textbf {\bibinfo {volume} {15}},\
  \bibinfo {pages} {5208} (\bibinfo {year} {2015})}\BibitemShut {NoStop}%
\bibitem [{\citenamefont {Majety}\ \emph {et~al.}(2023)\citenamefont {Majety},
  \citenamefont {Strohauer}, \citenamefont {Saha}, \citenamefont {Wietschorke},
  \citenamefont {Finley}, \citenamefont {M{\"u}ller},\ and\ \citenamefont
  {Radulaski}}]{Majety2023}%
  \BibitemOpen
  \bibfield  {author} {\bibinfo {author} {\bibfnamefont {S.}~\bibnamefont
  {Majety}}, \bibinfo {author} {\bibfnamefont {S.}~\bibnamefont {Strohauer}},
  \bibinfo {author} {\bibfnamefont {P.}~\bibnamefont {Saha}}, \bibinfo {author}
  {\bibfnamefont {F.}~\bibnamefont {Wietschorke}}, \bibinfo {author}
  {\bibfnamefont {J.~J.}\ \bibnamefont {Finley}}, \bibinfo {author}
  {\bibfnamefont {K.}~\bibnamefont {M{\"u}ller}},\ and\ \bibinfo {author}
  {\bibfnamefont {M.}~\bibnamefont {Radulaski}},\ }\bibfield  {title} {\bibinfo
  {title} {Triangular quantum photonic devices with integrated detectors in
  silicon carbide},\ }\href {https://doi.org/10.1088/2633-4356/acc302}
  {\bibfield  {journal} {\bibinfo  {journal} {Materials for Quantum
  Technology}\ }\textbf {\bibinfo {volume} {3}},\ \bibinfo {pages} {015004}
  (\bibinfo {year} {2023})}\BibitemShut {NoStop}%
\bibitem [{\citenamefont {Wollman}\ \emph {et~al.}(2021)\citenamefont
  {Wollman}, \citenamefont {Verma}, \citenamefont {Walter}, \citenamefont
  {Chiles}, \citenamefont {Korzh}, \citenamefont {Allmaras}, \citenamefont
  {Zhai}, \citenamefont {Lita}, \citenamefont {McCaughan}, \citenamefont
  {Schmidt}, \citenamefont {Frasca}, \citenamefont {Mirin}, \citenamefont
  {Nam},\ and\ \citenamefont {Shaw}}]{Wollman2021}%
  \BibitemOpen
  \bibfield  {author} {\bibinfo {author} {\bibfnamefont {E.~E.}\ \bibnamefont
  {Wollman}}, \bibinfo {author} {\bibfnamefont {V.~B.}\ \bibnamefont {Verma}},
  \bibinfo {author} {\bibfnamefont {A.~B.}\ \bibnamefont {Walter}}, \bibinfo
  {author} {\bibfnamefont {J.}~\bibnamefont {Chiles}}, \bibinfo {author}
  {\bibfnamefont {B.}~\bibnamefont {Korzh}}, \bibinfo {author} {\bibfnamefont
  {J.~P.}\ \bibnamefont {Allmaras}}, \bibinfo {author} {\bibfnamefont
  {Y.}~\bibnamefont {Zhai}}, \bibinfo {author} {\bibfnamefont {A.~E.}\
  \bibnamefont {Lita}}, \bibinfo {author} {\bibfnamefont {A.~N.}\ \bibnamefont
  {McCaughan}}, \bibinfo {author} {\bibfnamefont {E.}~\bibnamefont {Schmidt}},
  \bibinfo {author} {\bibfnamefont {S.}~\bibnamefont {Frasca}}, \bibinfo
  {author} {\bibfnamefont {R.~P.}\ \bibnamefont {Mirin}}, \bibinfo {author}
  {\bibfnamefont {S.~W.}\ \bibnamefont {Nam}},\ and\ \bibinfo {author}
  {\bibfnamefont {M.~D.}\ \bibnamefont {Shaw}},\ }\bibfield  {title} {\bibinfo
  {title} {Recent advances in superconducting nanowire single-photon detector
  technology for exoplanet transit spectroscopy in the mid-infrared},\ }\href
  {https://doi.org/10.1117/1.JATIS.7.1.011004} {\bibfield  {journal} {\bibinfo
  {journal} {Journal of Astronomical Telescopes, Instruments, and Systems}\
  }\textbf {\bibinfo {volume} {7}},\ \bibinfo {pages} {011004} (\bibinfo {year}
  {2021})}\BibitemShut {NoStop}%
\bibitem [{\citenamefont {Chiles}\ \emph {et~al.}(2022)\citenamefont {Chiles},
  \citenamefont {Charaev}, \citenamefont {Lasenby}, \citenamefont {Baryakhtar},
  \citenamefont {Huang}, \citenamefont {Roshko}, \citenamefont {Burton},
  \citenamefont {Colangelo}, \citenamefont {Van~Tilburg}, \citenamefont
  {Arvanitaki}, \citenamefont {Nam},\ and\ \citenamefont
  {Berggren}}]{Chiles2022}%
  \BibitemOpen
  \bibfield  {author} {\bibinfo {author} {\bibfnamefont {J.}~\bibnamefont
  {Chiles}}, \bibinfo {author} {\bibfnamefont {I.}~\bibnamefont {Charaev}},
  \bibinfo {author} {\bibfnamefont {R.}~\bibnamefont {Lasenby}}, \bibinfo
  {author} {\bibfnamefont {M.}~\bibnamefont {Baryakhtar}}, \bibinfo {author}
  {\bibfnamefont {J.}~\bibnamefont {Huang}}, \bibinfo {author} {\bibfnamefont
  {A.}~\bibnamefont {Roshko}}, \bibinfo {author} {\bibfnamefont
  {G.}~\bibnamefont {Burton}}, \bibinfo {author} {\bibfnamefont
  {M.}~\bibnamefont {Colangelo}}, \bibinfo {author} {\bibfnamefont
  {K.}~\bibnamefont {Van~Tilburg}}, \bibinfo {author} {\bibfnamefont
  {A.}~\bibnamefont {Arvanitaki}}, \bibinfo {author} {\bibfnamefont {S.~W.}\
  \bibnamefont {Nam}},\ and\ \bibinfo {author} {\bibfnamefont {K.~K.}\
  \bibnamefont {Berggren}},\ }\bibfield  {title} {\bibinfo {title} {New
  {{Constraints}} on {{Dark Photon Dark Matter}} with {{Superconducting
  Nanowire Detectors}} in an {{Optical Haloscope}}},\ }\href
  {https://doi.org/10.1103/PhysRevLett.128.231802} {\bibfield  {journal}
  {\bibinfo  {journal} {Physical Review Letters}\ }\textbf {\bibinfo {volume}
  {128}},\ \bibinfo {pages} {231802} (\bibinfo {year} {2022})}\BibitemShut
  {NoStop}%
\bibitem [{\citenamefont {Polakovic}\ \emph {et~al.}(2020)\citenamefont
  {Polakovic}, \citenamefont {Armstrong}, \citenamefont {Karapetrov},
  \citenamefont {Meziani},\ and\ \citenamefont {Novosad}}]{Polakovic2020}%
  \BibitemOpen
  \bibfield  {author} {\bibinfo {author} {\bibfnamefont {T.}~\bibnamefont
  {Polakovic}}, \bibinfo {author} {\bibfnamefont {W.}~\bibnamefont
  {Armstrong}}, \bibinfo {author} {\bibfnamefont {G.}~\bibnamefont
  {Karapetrov}}, \bibinfo {author} {\bibfnamefont {Z.~E.}\ \bibnamefont
  {Meziani}},\ and\ \bibinfo {author} {\bibfnamefont {V.}~\bibnamefont
  {Novosad}},\ }\bibfield  {title} {\bibinfo {title} {Unconventional
  applications of superconducting nanowire single photon detectors},\ }\href
  {https://doi.org/10.3390/nano10061198} {\bibfield  {journal} {\bibinfo
  {journal} {Nanomaterials}\ }\textbf {\bibinfo {volume} {10}},\ \bibinfo
  {pages} {1198} (\bibinfo {year} {2020})}\BibitemShut {NoStop}%
\bibitem [{\citenamefont {Shigefuji}\ \emph {et~al.}(2023)\citenamefont
  {Shigefuji}, \citenamefont {Osada}, \citenamefont {Yabuno}, \citenamefont
  {Miki}, \citenamefont {Terai},\ and\ \citenamefont
  {Noguchi}}]{Shigefuji2023}%
  \BibitemOpen
  \bibfield  {author} {\bibinfo {author} {\bibfnamefont {M.}~\bibnamefont
  {Shigefuji}}, \bibinfo {author} {\bibfnamefont {A.}~\bibnamefont {Osada}},
  \bibinfo {author} {\bibfnamefont {M.}~\bibnamefont {Yabuno}}, \bibinfo
  {author} {\bibfnamefont {S.}~\bibnamefont {Miki}}, \bibinfo {author}
  {\bibfnamefont {H.}~\bibnamefont {Terai}},\ and\ \bibinfo {author}
  {\bibfnamefont {A.}~\bibnamefont {Noguchi}},\ }\href@noop {} {\bibinfo
  {title} {Efficient low-energy single-electron detection using a large-area
  superconducting microstrip}} (\bibinfo {year} {2023}),\ \Eprint
  {https://arxiv.org/abs/2301.11212} {arxiv:2301.11212 [cond-mat,
  physics:physics, physics:quant-ph]} \BibitemShut {NoStop}%
\bibitem [{\citenamefont {Wollman}\ \emph {et~al.}(2019)\citenamefont
  {Wollman}, \citenamefont {Verma}, \citenamefont {Lita}, \citenamefont {Farr},
  \citenamefont {Shaw}, \citenamefont {Mirin},\ and\ \citenamefont
  {Woo~Nam}}]{Wollman2019}%
  \BibitemOpen
  \bibfield  {author} {\bibinfo {author} {\bibfnamefont {E.~E.}\ \bibnamefont
  {Wollman}}, \bibinfo {author} {\bibfnamefont {V.~B.}\ \bibnamefont {Verma}},
  \bibinfo {author} {\bibfnamefont {A.~E.}\ \bibnamefont {Lita}}, \bibinfo
  {author} {\bibfnamefont {W.~H.}\ \bibnamefont {Farr}}, \bibinfo {author}
  {\bibfnamefont {M.~D.}\ \bibnamefont {Shaw}}, \bibinfo {author}
  {\bibfnamefont {R.~P.}\ \bibnamefont {Mirin}},\ and\ \bibinfo {author}
  {\bibfnamefont {S.}~\bibnamefont {Woo~Nam}},\ }\bibfield  {title} {\bibinfo
  {title} {Kilopixel array of superconducting nanowire single-photon
  detectors},\ }\href {https://doi.org/10.1364/OE.27.035279} {\bibfield
  {journal} {\bibinfo  {journal} {Optics Express}\ }\textbf {\bibinfo {volume}
  {27}},\ \bibinfo {pages} {35279} (\bibinfo {year} {2019})}\BibitemShut
  {NoStop}%
\bibitem [{\citenamefont {McCaughan}\ \emph {et~al.}(2022)\citenamefont
  {McCaughan}, \citenamefont {Zhai}, \citenamefont {Korzh}, \citenamefont
  {Allmaras}, \citenamefont {Oripov}, \citenamefont {Shaw},\ and\ \citenamefont
  {Nam}}]{McCaughan2022}%
  \BibitemOpen
  \bibfield  {author} {\bibinfo {author} {\bibfnamefont {A.~N.}\ \bibnamefont
  {McCaughan}}, \bibinfo {author} {\bibfnamefont {Y.}~\bibnamefont {Zhai}},
  \bibinfo {author} {\bibfnamefont {B.}~\bibnamefont {Korzh}}, \bibinfo
  {author} {\bibfnamefont {J.~P.}\ \bibnamefont {Allmaras}}, \bibinfo {author}
  {\bibfnamefont {B.~G.}\ \bibnamefont {Oripov}}, \bibinfo {author}
  {\bibfnamefont {M.~D.}\ \bibnamefont {Shaw}},\ and\ \bibinfo {author}
  {\bibfnamefont {S.~W.}\ \bibnamefont {Nam}},\ }\bibfield  {title} {\bibinfo
  {title} {The thermally coupled imager: {{A}} scalable readout architecture
  for superconducting nanowire single photon detectors},\ }\href
  {https://doi.org/10.1063/5.0102154} {\bibfield  {journal} {\bibinfo
  {journal} {Applied Physics Letters}\ }\textbf {\bibinfo {volume} {121}},\
  \bibinfo {pages} {102602} (\bibinfo {year} {2022})}\BibitemShut {NoStop}%
\bibitem [{\citenamefont {Hortensius}\ \emph {et~al.}(2013)\citenamefont
  {Hortensius}, \citenamefont {Driessen},\ and\ \citenamefont
  {Klapwijk}}]{Hortensius2013}%
  \BibitemOpen
  \bibfield  {author} {\bibinfo {author} {\bibfnamefont {H.~L.}\ \bibnamefont
  {Hortensius}}, \bibinfo {author} {\bibfnamefont {E.~F.~C.}\ \bibnamefont
  {Driessen}},\ and\ \bibinfo {author} {\bibfnamefont {T.~M.}\ \bibnamefont
  {Klapwijk}},\ }\bibfield  {title} {\bibinfo {title} {Possible {{Indications}}
  of {{Electronic Inhomogeneities}} in {{Superconducting Nanowire
  Detectors}}},\ }\href {https://doi.org/10.1109/TASC.2013.2237935} {\bibfield
  {journal} {\bibinfo  {journal} {IEEE Transactions on Applied
  Superconductivity}\ }\textbf {\bibinfo {volume} {23}},\ \bibinfo {pages}
  {2200705} (\bibinfo {year} {2013})}\BibitemShut {NoStop}%
\bibitem [{\citenamefont {Noat}\ \emph {et~al.}(2013)\citenamefont {Noat},
  \citenamefont {Cherkez}, \citenamefont {Brun}, \citenamefont {Cren},
  \citenamefont {Carbillet}, \citenamefont {Debontridder}, \citenamefont
  {Ilin}, \citenamefont {Siegel}, \citenamefont {Semenov}, \citenamefont
  {H{\"u}bers},\ and\ \citenamefont {Roditchev}}]{Noat2013}%
  \BibitemOpen
  \bibfield  {author} {\bibinfo {author} {\bibfnamefont {Y.}~\bibnamefont
  {Noat}}, \bibinfo {author} {\bibfnamefont {V.}~\bibnamefont {Cherkez}},
  \bibinfo {author} {\bibfnamefont {C.}~\bibnamefont {Brun}}, \bibinfo {author}
  {\bibfnamefont {T.}~\bibnamefont {Cren}}, \bibinfo {author} {\bibfnamefont
  {C.}~\bibnamefont {Carbillet}}, \bibinfo {author} {\bibfnamefont
  {F.}~\bibnamefont {Debontridder}}, \bibinfo {author} {\bibfnamefont
  {K.}~\bibnamefont {Ilin}}, \bibinfo {author} {\bibfnamefont {M.}~\bibnamefont
  {Siegel}}, \bibinfo {author} {\bibfnamefont {A.}~\bibnamefont {Semenov}},
  \bibinfo {author} {\bibfnamefont {H.-W.}\ \bibnamefont {H{\"u}bers}},\ and\
  \bibinfo {author} {\bibfnamefont {D.}~\bibnamefont {Roditchev}},\ }\bibfield
  {title} {\bibinfo {title} {Unconventional superconductivity in ultrathin
  superconducting {{NbN}} films studied by scanning tunneling spectroscopy},\
  }\href {https://doi.org/10.1103/PhysRevB.88.014503} {\bibfield  {journal}
  {\bibinfo  {journal} {Physical Review B}\ }\textbf {\bibinfo {volume} {88}},\
  \bibinfo {pages} {014503} (\bibinfo {year} {2013})}\BibitemShut {NoStop}%
\bibitem [{\citenamefont {Sac{\'e}p{\'e}}\ \emph {et~al.}(2008)\citenamefont
  {Sac{\'e}p{\'e}}, \citenamefont {Chapelier}, \citenamefont {Baturina},
  \citenamefont {Vinokur}, \citenamefont {Baklanov},\ and\ \citenamefont
  {Sanquer}}]{Sacepe2008}%
  \BibitemOpen
  \bibfield  {author} {\bibinfo {author} {\bibfnamefont {B.}~\bibnamefont
  {Sac{\'e}p{\'e}}}, \bibinfo {author} {\bibfnamefont {C.}~\bibnamefont
  {Chapelier}}, \bibinfo {author} {\bibfnamefont {T.~I.}\ \bibnamefont
  {Baturina}}, \bibinfo {author} {\bibfnamefont {V.~M.}\ \bibnamefont
  {Vinokur}}, \bibinfo {author} {\bibfnamefont {M.~R.}\ \bibnamefont
  {Baklanov}},\ and\ \bibinfo {author} {\bibfnamefont {M.}~\bibnamefont
  {Sanquer}},\ }\bibfield  {title} {\bibinfo {title} {Disorder-{{Induced
  Inhomogeneities}} of the {{Superconducting State Close}} to the
  {{Superconductor-Insulator Transition}}},\ }\href
  {https://doi.org/10.1103/PhysRevLett.101.157006} {\bibfield  {journal}
  {\bibinfo  {journal} {Physical Review Letters}\ }\textbf {\bibinfo {volume}
  {101}},\ \bibinfo {pages} {157006} (\bibinfo {year} {2008})}\BibitemShut
  {NoStop}%
\bibitem [{\citenamefont {Kirtley}\ \emph {et~al.}(1987)\citenamefont
  {Kirtley}, \citenamefont {Raider}, \citenamefont {Feenstra},\ and\
  \citenamefont {Fein}}]{Kirtley1987}%
  \BibitemOpen
  \bibfield  {author} {\bibinfo {author} {\bibfnamefont {J.~R.}\ \bibnamefont
  {Kirtley}}, \bibinfo {author} {\bibfnamefont {S.~I.}\ \bibnamefont {Raider}},
  \bibinfo {author} {\bibfnamefont {R.~M.}\ \bibnamefont {Feenstra}},\ and\
  \bibinfo {author} {\bibfnamefont {A.~P.}\ \bibnamefont {Fein}},\ }\bibfield
  {title} {\bibinfo {title} {Spatial variation of the observed energy gap in
  granular superconducting {{NbN}} films},\ }\href
  {https://doi.org/10.1063/1.97795} {\bibfield  {journal} {\bibinfo  {journal}
  {Applied Physics Letters}\ }\textbf {\bibinfo {volume} {50}},\ \bibinfo
  {pages} {1607} (\bibinfo {year} {1987})}\BibitemShut {NoStop}%
\bibitem [{\citenamefont {Allman}\ \emph {et~al.}(2015)\citenamefont {Allman},
  \citenamefont {Verma}, \citenamefont {Stevens}, \citenamefont {Gerrits},
  \citenamefont {Horansky}, \citenamefont {Lita}, \citenamefont {Marsili},
  \citenamefont {Beyer}, \citenamefont {Shaw}, \citenamefont {Kumor},
  \citenamefont {Mirin},\ and\ \citenamefont {Nam}}]{Allman2015}%
  \BibitemOpen
  \bibfield  {author} {\bibinfo {author} {\bibfnamefont {M.~S.}\ \bibnamefont
  {Allman}}, \bibinfo {author} {\bibfnamefont {V.~B.}\ \bibnamefont {Verma}},
  \bibinfo {author} {\bibfnamefont {M.}~\bibnamefont {Stevens}}, \bibinfo
  {author} {\bibfnamefont {T.}~\bibnamefont {Gerrits}}, \bibinfo {author}
  {\bibfnamefont {R.~D.}\ \bibnamefont {Horansky}}, \bibinfo {author}
  {\bibfnamefont {A.~E.}\ \bibnamefont {Lita}}, \bibinfo {author}
  {\bibfnamefont {F.}~\bibnamefont {Marsili}}, \bibinfo {author} {\bibfnamefont
  {A.}~\bibnamefont {Beyer}}, \bibinfo {author} {\bibfnamefont {M.~D.}\
  \bibnamefont {Shaw}}, \bibinfo {author} {\bibfnamefont {D.}~\bibnamefont
  {Kumor}}, \bibinfo {author} {\bibfnamefont {R.}~\bibnamefont {Mirin}},\ and\
  \bibinfo {author} {\bibfnamefont {S.~W.}\ \bibnamefont {Nam}},\ }\bibfield
  {title} {\bibinfo {title} {A near-infrared 64-pixel superconducting nanowire
  single photon detector array with integrated multiplexed readout},\ }\href
  {https://doi.org/10.1063/1.4921318} {\bibfield  {journal} {\bibinfo
  {journal} {Applied Physics Letters}\ }\textbf {\bibinfo {volume} {106}},\
  \bibinfo {pages} {192601} (\bibinfo {year} {2015})}\BibitemShut {NoStop}%
\bibitem [{\citenamefont {Gaudio}\ \emph {et~al.}(2014)\citenamefont {Gaudio},
  \citenamefont {{op 't Hoog}}, \citenamefont {Zhou}, \citenamefont {Sahin},\
  and\ \citenamefont {Fiore}}]{Gaudio2014}%
  \BibitemOpen
  \bibfield  {author} {\bibinfo {author} {\bibfnamefont {R.}~\bibnamefont
  {Gaudio}}, \bibinfo {author} {\bibfnamefont {K.~P.~M.}\ \bibnamefont {{op 't
  Hoog}}}, \bibinfo {author} {\bibfnamefont {Z.}~\bibnamefont {Zhou}}, \bibinfo
  {author} {\bibfnamefont {D.}~\bibnamefont {Sahin}},\ and\ \bibinfo {author}
  {\bibfnamefont {A.}~\bibnamefont {Fiore}},\ }\bibfield  {title} {\bibinfo
  {title} {Inhomogeneous critical current in nanowire superconducting
  single-photon detectors},\ }\href {https://doi.org/10.1063/1.4903071}
  {\bibfield  {journal} {\bibinfo  {journal} {Applied Physics Letters}\
  }\textbf {\bibinfo {volume} {105}},\ \bibinfo {pages} {222602} (\bibinfo
  {year} {2014})}\BibitemShut {NoStop}%
\bibitem [{\citenamefont {Kerman}\ \emph {et~al.}(2007)\citenamefont {Kerman},
  \citenamefont {Dauler}, \citenamefont {Yang}, \citenamefont {Rosfjord},
  \citenamefont {Anant}, \citenamefont {Berggren}, \citenamefont {Gol'tsman},\
  and\ \citenamefont {Voronov}}]{Kerman2007}%
  \BibitemOpen
  \bibfield  {author} {\bibinfo {author} {\bibfnamefont {A.~J.}\ \bibnamefont
  {Kerman}}, \bibinfo {author} {\bibfnamefont {E.~A.}\ \bibnamefont {Dauler}},
  \bibinfo {author} {\bibfnamefont {J.~K.}\ \bibnamefont {Yang}}, \bibinfo
  {author} {\bibfnamefont {K.~M.}\ \bibnamefont {Rosfjord}}, \bibinfo {author}
  {\bibfnamefont {V.}~\bibnamefont {Anant}}, \bibinfo {author} {\bibfnamefont
  {K.~K.}\ \bibnamefont {Berggren}}, \bibinfo {author} {\bibfnamefont {G.~N.}\
  \bibnamefont {Gol'tsman}},\ and\ \bibinfo {author} {\bibfnamefont {B.~M.}\
  \bibnamefont {Voronov}},\ }\bibfield  {title} {\bibinfo {title}
  {Constriction-limited detection efficiency of superconducting nanowire
  single-photon detectors},\ }\href {https://doi.org/10.1063/1.2696926}
  {\bibfield  {journal} {\bibinfo  {journal} {Applied Physics Letters}\
  }\textbf {\bibinfo {volume} {90}},\ \bibinfo {pages} {101110} (\bibinfo
  {year} {2007})}\BibitemShut {NoStop}%
\bibitem [{\citenamefont {Esmaeil~Zadeh}\ \emph {et~al.}(2021)\citenamefont
  {Esmaeil~Zadeh}, \citenamefont {Chang}, \citenamefont {Los}, \citenamefont
  {Gyger}, \citenamefont {Elshaari}, \citenamefont {Steinhauer}, \citenamefont
  {Dorenbos},\ and\ \citenamefont {Zwiller}}]{EsmaeilZadeh2021}%
  \BibitemOpen
  \bibfield  {author} {\bibinfo {author} {\bibfnamefont {I.}~\bibnamefont
  {Esmaeil~Zadeh}}, \bibinfo {author} {\bibfnamefont {J.}~\bibnamefont
  {Chang}}, \bibinfo {author} {\bibfnamefont {J.~W.}\ \bibnamefont {Los}},
  \bibinfo {author} {\bibfnamefont {S.}~\bibnamefont {Gyger}}, \bibinfo
  {author} {\bibfnamefont {A.~W.}\ \bibnamefont {Elshaari}}, \bibinfo {author}
  {\bibfnamefont {S.}~\bibnamefont {Steinhauer}}, \bibinfo {author}
  {\bibfnamefont {S.~N.}\ \bibnamefont {Dorenbos}},\ and\ \bibinfo {author}
  {\bibfnamefont {V.}~\bibnamefont {Zwiller}},\ }\bibfield  {title} {\bibinfo
  {title} {Superconducting nanowire single-photon detectors: {{A}} perspective
  on evolution, state-of-the-art, future developments, and applications},\
  }\href {https://doi.org/10.1063/5.0045990} {\bibfield  {journal} {\bibinfo
  {journal} {Applied Physics Letters}\ }\textbf {\bibinfo {volume} {118}},\
  \bibinfo {pages} {190502} (\bibinfo {year} {2021})}\BibitemShut {NoStop}%
\bibitem [{\citenamefont {Marsili}\ \emph {et~al.}(2013)\citenamefont
  {Marsili}, \citenamefont {Verma}, \citenamefont {Stern}, \citenamefont
  {Harrington}, \citenamefont {Lita}, \citenamefont {Gerrits}, \citenamefont
  {Vayshenker}, \citenamefont {Baek}, \citenamefont {Shaw}, \citenamefont
  {Mirin},\ and\ \citenamefont {Nam}}]{Marsili2013}%
  \BibitemOpen
  \bibfield  {author} {\bibinfo {author} {\bibfnamefont {F.}~\bibnamefont
  {Marsili}}, \bibinfo {author} {\bibfnamefont {V.~B.}\ \bibnamefont {Verma}},
  \bibinfo {author} {\bibfnamefont {J.~A.}\ \bibnamefont {Stern}}, \bibinfo
  {author} {\bibfnamefont {S.}~\bibnamefont {Harrington}}, \bibinfo {author}
  {\bibfnamefont {A.~E.}\ \bibnamefont {Lita}}, \bibinfo {author}
  {\bibfnamefont {T.}~\bibnamefont {Gerrits}}, \bibinfo {author} {\bibfnamefont
  {I.}~\bibnamefont {Vayshenker}}, \bibinfo {author} {\bibfnamefont
  {B.}~\bibnamefont {Baek}}, \bibinfo {author} {\bibfnamefont {M.~D.}\
  \bibnamefont {Shaw}}, \bibinfo {author} {\bibfnamefont {R.~P.}\ \bibnamefont
  {Mirin}},\ and\ \bibinfo {author} {\bibfnamefont {S.~W.}\ \bibnamefont
  {Nam}},\ }\bibfield  {title} {\bibinfo {title} {Detecting single infrared
  photons with 93\% system efficiency},\ }\href
  {https://doi.org/10.1038/nphoton.2013.13} {\bibfield  {journal} {\bibinfo
  {journal} {Nature Photonics}\ }\textbf {\bibinfo {volume} {7}},\ \bibinfo
  {pages} {210} (\bibinfo {year} {2013})}\BibitemShut {NoStop}%
\bibitem [{\citenamefont {Steinhauer}\ \emph {et~al.}(2021)\citenamefont
  {Steinhauer}, \citenamefont {Gyger},\ and\ \citenamefont
  {Zwiller}}]{Steinhauer2021a}%
  \BibitemOpen
  \bibfield  {author} {\bibinfo {author} {\bibfnamefont {S.}~\bibnamefont
  {Steinhauer}}, \bibinfo {author} {\bibfnamefont {S.}~\bibnamefont {Gyger}},\
  and\ \bibinfo {author} {\bibfnamefont {V.}~\bibnamefont {Zwiller}},\
  }\bibfield  {title} {\bibinfo {title} {Progress on large-scale
  superconducting nanowire single-photon detectors},\ }\href
  {https://doi.org/10.1063/5.0044057} {\bibfield  {journal} {\bibinfo
  {journal} {Applied Physics Letters}\ }\textbf {\bibinfo {volume} {118}},\
  \bibinfo {pages} {100501} (\bibinfo {year} {2021})}\BibitemShut {NoStop}%
\bibitem [{\citenamefont {Cheng}\ \emph {et~al.}(2019)\citenamefont {Cheng},
  \citenamefont {Wang},\ and\ \citenamefont {Tang}}]{Cheng2019a}%
  \BibitemOpen
  \bibfield  {author} {\bibinfo {author} {\bibfnamefont {R.}~\bibnamefont
  {Cheng}}, \bibinfo {author} {\bibfnamefont {S.}~\bibnamefont {Wang}},\ and\
  \bibinfo {author} {\bibfnamefont {H.~X.}\ \bibnamefont {Tang}},\ }\bibfield
  {title} {\bibinfo {title} {Superconducting nanowire single-photon detectors
  fabricated from atomic-layer-deposited {{NbN}}},\ }\href
  {https://doi.org/10.1063/1.5131664} {\bibfield  {journal} {\bibinfo
  {journal} {Applied Physics Letters}\ }\textbf {\bibinfo {volume} {115}},\
  \bibinfo {pages} {241101} (\bibinfo {year} {2019})}\BibitemShut {NoStop}%
\bibitem [{\citenamefont {Knehr}\ \emph {et~al.}(2019)\citenamefont {Knehr},
  \citenamefont {Kuzmin}, \citenamefont {Vodolazov}, \citenamefont {Ziegler},
  \citenamefont {Doerner}, \citenamefont {Ilin}, \citenamefont {Siegel},
  \citenamefont {Stolz},\ and\ \citenamefont {Schmidt}}]{Knehr2019}%
  \BibitemOpen
  \bibfield  {author} {\bibinfo {author} {\bibfnamefont {E.}~\bibnamefont
  {Knehr}}, \bibinfo {author} {\bibfnamefont {A.}~\bibnamefont {Kuzmin}},
  \bibinfo {author} {\bibfnamefont {D.~Y.}\ \bibnamefont {Vodolazov}}, \bibinfo
  {author} {\bibfnamefont {M.}~\bibnamefont {Ziegler}}, \bibinfo {author}
  {\bibfnamefont {S.}~\bibnamefont {Doerner}}, \bibinfo {author} {\bibfnamefont
  {K.}~\bibnamefont {Ilin}}, \bibinfo {author} {\bibfnamefont {M.}~\bibnamefont
  {Siegel}}, \bibinfo {author} {\bibfnamefont {R.}~\bibnamefont {Stolz}},\ and\
  \bibinfo {author} {\bibfnamefont {H.}~\bibnamefont {Schmidt}},\ }\bibfield
  {title} {\bibinfo {title} {Nanowire single-photon detectors made of atomic
  layer-deposited niobium nitride},\ }\href
  {https://doi.org/10.1088/1361-6668/ab48d7} {\bibfield  {journal} {\bibinfo
  {journal} {Superconductor Science and Technology}\ }\textbf {\bibinfo
  {volume} {32}},\ \bibinfo {pages} {125007} (\bibinfo {year}
  {2019})}\BibitemShut {NoStop}%
\bibitem [{\citenamefont {Cheng}\ \emph {et~al.}(2020)\citenamefont {Cheng},
  \citenamefont {Wright}, \citenamefont {Xing}, \citenamefont {Jena},\ and\
  \citenamefont {Tang}}]{Cheng2020}%
  \BibitemOpen
  \bibfield  {author} {\bibinfo {author} {\bibfnamefont {R.}~\bibnamefont
  {Cheng}}, \bibinfo {author} {\bibfnamefont {J.}~\bibnamefont {Wright}},
  \bibinfo {author} {\bibfnamefont {H.~G.}\ \bibnamefont {Xing}}, \bibinfo
  {author} {\bibfnamefont {D.}~\bibnamefont {Jena}},\ and\ \bibinfo {author}
  {\bibfnamefont {H.~X.}\ \bibnamefont {Tang}},\ }\bibfield  {title} {\bibinfo
  {title} {Epitaxial niobium nitride superconducting nanowire single-photon
  detectors},\ }\href {https://doi.org/10.1063/5.0018818} {\bibfield  {journal}
  {\bibinfo  {journal} {Applied Physics Letters}\ }\textbf {\bibinfo {volume}
  {117}},\ \bibinfo {pages} {132601} (\bibinfo {year} {2020})}\BibitemShut
  {NoStop}%
\bibitem [{\citenamefont {Zhang}\ \emph {et~al.}(2019)\citenamefont {Zhang},
  \citenamefont {Jia}, \citenamefont {You}, \citenamefont {Ou}, \citenamefont
  {Huang}, \citenamefont {Zhang}, \citenamefont {Li}, \citenamefont {Wang},\
  and\ \citenamefont {Xie}}]{Zhang2019}%
  \BibitemOpen
  \bibfield  {author} {\bibinfo {author} {\bibfnamefont {W.}~\bibnamefont
  {Zhang}}, \bibinfo {author} {\bibfnamefont {Q.}~\bibnamefont {Jia}}, \bibinfo
  {author} {\bibfnamefont {L.}~\bibnamefont {You}}, \bibinfo {author}
  {\bibfnamefont {X.}~\bibnamefont {Ou}}, \bibinfo {author} {\bibfnamefont
  {H.}~\bibnamefont {Huang}}, \bibinfo {author} {\bibfnamefont
  {L.}~\bibnamefont {Zhang}}, \bibinfo {author} {\bibfnamefont
  {H.}~\bibnamefont {Li}}, \bibinfo {author} {\bibfnamefont {Z.}~\bibnamefont
  {Wang}},\ and\ \bibinfo {author} {\bibfnamefont {X.}~\bibnamefont {Xie}},\
  }\bibfield  {title} {\bibinfo {title} {Saturating {{Intrinsic Detection
  Efficiency}} of {{Superconducting Nanowire Single-Photon Detectors}} via
  {{Defect Engineering}}},\ }\href
  {https://doi.org/10.1103/PhysRevApplied.12.044040} {\bibfield  {journal}
  {\bibinfo  {journal} {Physical Review Applied}\ }\textbf {\bibinfo {volume}
  {12}},\ \bibinfo {pages} {044040} (\bibinfo {year} {2019})}\BibitemShut
  {NoStop}%
\bibitem [{\citenamefont {Xu}\ \emph {et~al.}(2021)\citenamefont {Xu},
  \citenamefont {Zhang}, \citenamefont {You}, \citenamefont {Xiong},
  \citenamefont {Sun}, \citenamefont {Huang}, \citenamefont {Ou}, \citenamefont
  {Pan}, \citenamefont {Lv}, \citenamefont {Li}, \citenamefont {Wang},\ and\
  \citenamefont {Xie}}]{Xu2021}%
  \BibitemOpen
  \bibfield  {author} {\bibinfo {author} {\bibfnamefont {G.-Z.}\ \bibnamefont
  {Xu}}, \bibinfo {author} {\bibfnamefont {W.-J.}\ \bibnamefont {Zhang}},
  \bibinfo {author} {\bibfnamefont {L.-X.}\ \bibnamefont {You}}, \bibinfo
  {author} {\bibfnamefont {J.-M.}\ \bibnamefont {Xiong}}, \bibinfo {author}
  {\bibfnamefont {X.-Q.}\ \bibnamefont {Sun}}, \bibinfo {author} {\bibfnamefont
  {H.}~\bibnamefont {Huang}}, \bibinfo {author} {\bibfnamefont
  {X.}~\bibnamefont {Ou}}, \bibinfo {author} {\bibfnamefont {Y.-M.}\
  \bibnamefont {Pan}}, \bibinfo {author} {\bibfnamefont {C.-L.}\ \bibnamefont
  {Lv}}, \bibinfo {author} {\bibfnamefont {H.}~\bibnamefont {Li}}, \bibinfo
  {author} {\bibfnamefont {Z.}~\bibnamefont {Wang}},\ and\ \bibinfo {author}
  {\bibfnamefont {X.-M.}\ \bibnamefont {Xie}},\ }\bibfield  {title} {\bibinfo
  {title} {Superconducting microstrip single-photon detector with system
  detection efficiency over 90\% at 1550 nm},\ }\href
  {https://doi.org/10.1364/prj.419514} {\bibfield  {journal} {\bibinfo
  {journal} {Photonics Research}\ }\textbf {\bibinfo {volume} {9}},\ \bibinfo
  {pages} {958} (\bibinfo {year} {2021})}\BibitemShut {NoStop}%
\bibitem [{\citenamefont {Zhang}\ \emph {et~al.}(2021)\citenamefont {Zhang},
  \citenamefont {Xu}, \citenamefont {You}, \citenamefont {Zhang}, \citenamefont
  {Huang}, \citenamefont {Ou}, \citenamefont {Sun}, \citenamefont {Xiong},
  \citenamefont {Li}, \citenamefont {Wang},\ and\ \citenamefont
  {Xie}}]{Zhang2021}%
  \BibitemOpen
  \bibfield  {author} {\bibinfo {author} {\bibfnamefont {W.-J.}\ \bibnamefont
  {Zhang}}, \bibinfo {author} {\bibfnamefont {G.-Z.}\ \bibnamefont {Xu}},
  \bibinfo {author} {\bibfnamefont {L.-X.}\ \bibnamefont {You}}, \bibinfo
  {author} {\bibfnamefont {C.-J.}\ \bibnamefont {Zhang}}, \bibinfo {author}
  {\bibfnamefont {H.}~\bibnamefont {Huang}}, \bibinfo {author} {\bibfnamefont
  {X.}~\bibnamefont {Ou}}, \bibinfo {author} {\bibfnamefont {X.-Q.}\
  \bibnamefont {Sun}}, \bibinfo {author} {\bibfnamefont {J.-M.}\ \bibnamefont
  {Xiong}}, \bibinfo {author} {\bibfnamefont {H.}~\bibnamefont {Li}}, \bibinfo
  {author} {\bibfnamefont {Z.}~\bibnamefont {Wang}},\ and\ \bibinfo {author}
  {\bibfnamefont {X.-M.}\ \bibnamefont {Xie}},\ }\bibfield  {title} {\bibinfo
  {title} {Sixteen-channel fiber array-coupled superconducting single-photon
  detector array with average system detection efficiency over 60\% at telecom
  wavelength},\ }\href {https://doi.org/10.1364/OL.418219} {\bibfield
  {journal} {\bibinfo  {journal} {Optics Letters}\ }\textbf {\bibinfo {volume}
  {46}},\ \bibinfo {pages} {1049} (\bibinfo {year} {2021})}\BibitemShut
  {NoStop}%
\bibitem [{\citenamefont {Zhang}\ \emph {et~al.}(2022)\citenamefont {Zhang},
  \citenamefont {Zhang}, \citenamefont {Zhou}, \citenamefont {Zhang},
  \citenamefont {You}, \citenamefont {Li}, \citenamefont {Fan}, \citenamefont
  {Pan}, \citenamefont {Yu}, \citenamefont {Li},\ and\ \citenamefont
  {Wang}}]{Zhang2022}%
  \BibitemOpen
  \bibfield  {author} {\bibinfo {author} {\bibfnamefont {X.}~\bibnamefont
  {Zhang}}, \bibinfo {author} {\bibfnamefont {W.}~\bibnamefont {Zhang}},
  \bibinfo {author} {\bibfnamefont {H.}~\bibnamefont {Zhou}}, \bibinfo {author}
  {\bibfnamefont {X.}~\bibnamefont {Zhang}}, \bibinfo {author} {\bibfnamefont
  {L.}~\bibnamefont {You}}, \bibinfo {author} {\bibfnamefont {H.}~\bibnamefont
  {Li}}, \bibinfo {author} {\bibfnamefont {D.}~\bibnamefont {Fan}}, \bibinfo
  {author} {\bibfnamefont {Y.}~\bibnamefont {Pan}}, \bibinfo {author}
  {\bibfnamefont {H.}~\bibnamefont {Yu}}, \bibinfo {author} {\bibfnamefont
  {L.}~\bibnamefont {Li}},\ and\ \bibinfo {author} {\bibfnamefont
  {Z.}~\bibnamefont {Wang}},\ }\bibfield  {title} {\bibinfo {title} {{{NbN
  Superconducting Nanowire Single-Photon Detector}} with 90.5\% {{Saturated
  System Detection Efficiency}} and 14.7 ps {{System Jitter}} at 1550 nm
  {{Wavelength}}},\ }\href {https://doi.org/10.1109/JSTQE.2022.3153029}
  {\bibfield  {journal} {\bibinfo  {journal} {IEEE Journal of Selected Topics
  in Quantum Electronics}\ }\textbf {\bibinfo {volume} {28}},\ \bibinfo {pages}
  {1} (\bibinfo {year} {2022})}\BibitemShut {NoStop}%
\bibitem [{\citenamefont {Flaschmann}\ \emph {et~al.}(2023)\citenamefont
  {Flaschmann}, \citenamefont {Zugliani}, \citenamefont {Schmid}, \citenamefont
  {Spedicato}, \citenamefont {Strohauer}, \citenamefont {Wietschorke},
  \citenamefont {Flassig}, \citenamefont {Finley},\ and\ \citenamefont
  {M{\"u}ller}}]{Flaschmann2023}%
  \BibitemOpen
  \bibfield  {author} {\bibinfo {author} {\bibfnamefont {R.}~\bibnamefont
  {Flaschmann}}, \bibinfo {author} {\bibfnamefont {L.}~\bibnamefont
  {Zugliani}}, \bibinfo {author} {\bibfnamefont {C.}~\bibnamefont {Schmid}},
  \bibinfo {author} {\bibfnamefont {S.}~\bibnamefont {Spedicato}}, \bibinfo
  {author} {\bibfnamefont {S.}~\bibnamefont {Strohauer}}, \bibinfo {author}
  {\bibfnamefont {F.}~\bibnamefont {Wietschorke}}, \bibinfo {author}
  {\bibfnamefont {F.}~\bibnamefont {Flassig}}, \bibinfo {author} {\bibfnamefont
  {J.~J.}\ \bibnamefont {Finley}},\ and\ \bibinfo {author} {\bibfnamefont
  {K.}~\bibnamefont {M{\"u}ller}},\ }\bibfield  {title} {\bibinfo {title} {The
  dependence of timing jitter of superconducting nanowire single-photon
  detectors on the multi-layer sample design and slew rate},\ }\href
  {https://doi.org/10.1039/D2NR04494C} {\bibfield  {journal} {\bibinfo
  {journal} {Nanoscale}\ }\textbf {\bibinfo {volume} {15}},\ \bibinfo {pages}
  {1086} (\bibinfo {year} {2023})}\BibitemShut {NoStop}%
\bibitem [{\citenamefont {{van der Pauw}}(1958)}]{vanderPauw1958}%
  \BibitemOpen
  \bibfield  {author} {\bibinfo {author} {\bibfnamefont {L.~J.}\ \bibnamefont
  {{van der Pauw}}},\ }\bibfield  {title} {\bibinfo {title} {A method of
  measuring the resistivity and {{Hall}} coefficient on lamellae of arbitrary
  shape},\ }\href@noop {} {\bibfield  {journal} {\bibinfo  {journal} {Philips
  Technical Review}\ }\textbf {\bibinfo {volume} {20}},\ \bibinfo {pages} {220}
  (\bibinfo {year} {1958})}\BibitemShut {NoStop}%
\bibitem [{\citenamefont {Miccoli}\ \emph {et~al.}(2015)\citenamefont
  {Miccoli}, \citenamefont {Edler}, \citenamefont {Pfn{\"u}r},\ and\
  \citenamefont {Tegenkamp}}]{Miccoli2015}%
  \BibitemOpen
  \bibfield  {author} {\bibinfo {author} {\bibfnamefont {I.}~\bibnamefont
  {Miccoli}}, \bibinfo {author} {\bibfnamefont {F.}~\bibnamefont {Edler}},
  \bibinfo {author} {\bibfnamefont {H.}~\bibnamefont {Pfn{\"u}r}},\ and\
  \bibinfo {author} {\bibfnamefont {C.}~\bibnamefont {Tegenkamp}},\ }\bibfield
  {title} {\bibinfo {title} {The 100th anniversary of the four-point probe
  technique: The role of probe geometries in isotropic and anisotropic
  systems},\ }\href {https://doi.org/10.1088/0953-8984/27/22/223201} {\bibfield
   {journal} {\bibinfo  {journal} {Journal of Physics: Condensed Matter}\
  }\textbf {\bibinfo {volume} {27}},\ \bibinfo {pages} {223201} (\bibinfo
  {year} {2015})}\BibitemShut {NoStop}%
\bibitem [{\citenamefont {Sidorova}\ \emph {et~al.}(2021)\citenamefont
  {Sidorova}, \citenamefont {Semenov}, \citenamefont {H{\"u}bers},
  \citenamefont {Gyger}, \citenamefont {Steinhauer}, \citenamefont {Zhang},\
  and\ \citenamefont {Schilling}}]{Sidorova2021}%
  \BibitemOpen
  \bibfield  {author} {\bibinfo {author} {\bibfnamefont {M.}~\bibnamefont
  {Sidorova}}, \bibinfo {author} {\bibfnamefont {A.~D.}\ \bibnamefont
  {Semenov}}, \bibinfo {author} {\bibfnamefont {H.~W.}\ \bibnamefont
  {H{\"u}bers}}, \bibinfo {author} {\bibfnamefont {S.}~\bibnamefont {Gyger}},
  \bibinfo {author} {\bibfnamefont {S.}~\bibnamefont {Steinhauer}}, \bibinfo
  {author} {\bibfnamefont {X.}~\bibnamefont {Zhang}},\ and\ \bibinfo {author}
  {\bibfnamefont {A.}~\bibnamefont {Schilling}},\ }\bibfield  {title} {\bibinfo
  {title} {Magnetoconductance and photoresponse properties of disordered
  {{NbTiN}} films},\ }\href {https://doi.org/10.1103/PhysRevB.104.184514}
  {\bibfield  {journal} {\bibinfo  {journal} {Physical Review B}\ }\textbf
  {\bibinfo {volume} {104}},\ \bibinfo {pages} {184514} (\bibinfo {year}
  {2021})}\BibitemShut {NoStop}%
\bibitem [{\citenamefont {Semenov}\ \emph {et~al.}(2009)\citenamefont
  {Semenov}, \citenamefont {G{\"u}nther}, \citenamefont {B{\"o}ttger},
  \citenamefont {H{\"u}bers}, \citenamefont {Bartolf}, \citenamefont {Engel},
  \citenamefont {Schilling}, \citenamefont {Ilin}, \citenamefont {Siegel},
  \citenamefont {Schneider}, \citenamefont {Gerthsen},\ and\ \citenamefont
  {Gippius}}]{Semenov2009}%
  \BibitemOpen
  \bibfield  {author} {\bibinfo {author} {\bibfnamefont {A.}~\bibnamefont
  {Semenov}}, \bibinfo {author} {\bibfnamefont {B.}~\bibnamefont
  {G{\"u}nther}}, \bibinfo {author} {\bibfnamefont {U.}~\bibnamefont
  {B{\"o}ttger}}, \bibinfo {author} {\bibfnamefont {H.-W.}\ \bibnamefont
  {H{\"u}bers}}, \bibinfo {author} {\bibfnamefont {H.}~\bibnamefont {Bartolf}},
  \bibinfo {author} {\bibfnamefont {A.}~\bibnamefont {Engel}}, \bibinfo
  {author} {\bibfnamefont {A.}~\bibnamefont {Schilling}}, \bibinfo {author}
  {\bibfnamefont {K.}~\bibnamefont {Ilin}}, \bibinfo {author} {\bibfnamefont
  {M.}~\bibnamefont {Siegel}}, \bibinfo {author} {\bibfnamefont
  {R.}~\bibnamefont {Schneider}}, \bibinfo {author} {\bibfnamefont
  {D.}~\bibnamefont {Gerthsen}},\ and\ \bibinfo {author} {\bibfnamefont
  {N.~A.}\ \bibnamefont {Gippius}},\ }\bibfield  {title} {\bibinfo {title}
  {Optical and transport properties of ultrathin {{NbN}} films and
  nanostructures},\ }\href {https://doi.org/10.1103/PhysRevB.80.054510}
  {\bibfield  {journal} {\bibinfo  {journal} {Physical Review B}\ }\textbf
  {\bibinfo {volume} {80}},\ \bibinfo {pages} {054510} (\bibinfo {year}
  {2009})}\BibitemShut {NoStop}%
\bibitem [{\citenamefont {Banerjee}\ \emph {et~al.}(2018)\citenamefont
  {Banerjee}, \citenamefont {Heath}, \citenamefont {Morozov}, \citenamefont
  {Hemakumara}, \citenamefont {Nasti}, \citenamefont {Thayne},\ and\
  \citenamefont {Hadfield}}]{Banerjee2018}%
  \BibitemOpen
  \bibfield  {author} {\bibinfo {author} {\bibfnamefont {A.}~\bibnamefont
  {Banerjee}}, \bibinfo {author} {\bibfnamefont {R.~M.}\ \bibnamefont {Heath}},
  \bibinfo {author} {\bibfnamefont {D.}~\bibnamefont {Morozov}}, \bibinfo
  {author} {\bibfnamefont {D.}~\bibnamefont {Hemakumara}}, \bibinfo {author}
  {\bibfnamefont {U.}~\bibnamefont {Nasti}}, \bibinfo {author} {\bibfnamefont
  {I.}~\bibnamefont {Thayne}},\ and\ \bibinfo {author} {\bibfnamefont {R.~H.}\
  \bibnamefont {Hadfield}},\ }\bibfield  {title} {\bibinfo {title} {Optical
  properties of refractory metal based thin films},\ }\href
  {https://doi.org/10.1364/ome.8.002072} {\bibfield  {journal} {\bibinfo
  {journal} {Optical Materials Express}\ }\textbf {\bibinfo {volume} {8}},\
  \bibinfo {pages} {2072} (\bibinfo {year} {2018})}\BibitemShut {NoStop}%
\bibitem [{\citenamefont {Zhang}\ \emph {et~al.}(2018)\citenamefont {Zhang},
  \citenamefont {You}, \citenamefont {Ying}, \citenamefont {Peng},\ and\
  \citenamefont {Wang}}]{Zhang2018b}%
  \BibitemOpen
  \bibfield  {author} {\bibinfo {author} {\bibfnamefont {L.}~\bibnamefont
  {Zhang}}, \bibinfo {author} {\bibfnamefont {L.}~\bibnamefont {You}}, \bibinfo
  {author} {\bibfnamefont {L.}~\bibnamefont {Ying}}, \bibinfo {author}
  {\bibfnamefont {W.}~\bibnamefont {Peng}},\ and\ \bibinfo {author}
  {\bibfnamefont {Z.}~\bibnamefont {Wang}},\ }\bibfield  {title} {\bibinfo
  {title} {Characterization of surface oxidation layers on ultrathin {{NbTiN}}
  films},\ }\href {https://doi.org/10.1016/j.physc.2017.10.008} {\bibfield
  {journal} {\bibinfo  {journal} {Physica C: Superconductivity and its
  Applications}\ }\textbf {\bibinfo {volume} {545}},\ \bibinfo {pages} {1}
  (\bibinfo {year} {2018})}\BibitemShut {NoStop}%
\bibitem [{\citenamefont {Korneeva}\ \emph {et~al.}(2018)\citenamefont
  {Korneeva}, \citenamefont {Vodolazov}, \citenamefont {Semenov}, \citenamefont
  {Florya}, \citenamefont {Simonov}, \citenamefont {Baeva}, \citenamefont
  {Korneev}, \citenamefont {Goltsman},\ and\ \citenamefont
  {Klapwijk}}]{Korneeva2018}%
  \BibitemOpen
  \bibfield  {author} {\bibinfo {author} {\bibfnamefont {Y.~P.}\ \bibnamefont
  {Korneeva}}, \bibinfo {author} {\bibfnamefont {D.~Y.}\ \bibnamefont
  {Vodolazov}}, \bibinfo {author} {\bibfnamefont {A.~V.}\ \bibnamefont
  {Semenov}}, \bibinfo {author} {\bibfnamefont {I.~N.}\ \bibnamefont {Florya}},
  \bibinfo {author} {\bibfnamefont {N.}~\bibnamefont {Simonov}}, \bibinfo
  {author} {\bibfnamefont {E.}~\bibnamefont {Baeva}}, \bibinfo {author}
  {\bibfnamefont {A.~A.}\ \bibnamefont {Korneev}}, \bibinfo {author}
  {\bibfnamefont {G.~N.}\ \bibnamefont {Goltsman}},\ and\ \bibinfo {author}
  {\bibfnamefont {T.~M.}\ \bibnamefont {Klapwijk}},\ }\bibfield  {title}
  {\bibinfo {title} {Optical {{Single-Photon Detection}} in {{Micrometer-Scale
  NbN Bridges}}},\ }\href {https://doi.org/10.1103/PhysRevApplied.9.064037}
  {\bibfield  {journal} {\bibinfo  {journal} {Physical Review Applied}\
  }\textbf {\bibinfo {volume} {9}},\ \bibinfo {pages} {64037} (\bibinfo {year}
  {2018})}\BibitemShut {NoStop}%
\bibitem [{\citenamefont {Clem}\ and\ \citenamefont {Kogan}(2012)}]{Clem2012}%
  \BibitemOpen
  \bibfield  {author} {\bibinfo {author} {\bibfnamefont {J.~R.}\ \bibnamefont
  {Clem}}\ and\ \bibinfo {author} {\bibfnamefont {V.~G.}\ \bibnamefont
  {Kogan}},\ }\bibfield  {title} {\bibinfo {title} {Kinetic impedance and
  depairing in thin and narrow superconducting films},\ }\href
  {https://doi.org/10.1103/PhysRevB.86.174521} {\bibfield  {journal} {\bibinfo
  {journal} {Physical Review B}\ }\textbf {\bibinfo {volume} {86}},\ \bibinfo
  {pages} {174521} (\bibinfo {year} {2012})}\BibitemShut {NoStop}%
\bibitem [{\citenamefont {Il'in}\ \emph {et~al.}(2008)\citenamefont {Il'in},
  \citenamefont {Siegel}, \citenamefont {Engel}, \citenamefont {Bartolf},
  \citenamefont {Schilling}, \citenamefont {Semenov},\ and\ \citenamefont
  {Huebers}}]{Ilin2008}%
  \BibitemOpen
  \bibfield  {author} {\bibinfo {author} {\bibfnamefont {K.}~\bibnamefont
  {Il'in}}, \bibinfo {author} {\bibfnamefont {M.}~\bibnamefont {Siegel}},
  \bibinfo {author} {\bibfnamefont {A.}~\bibnamefont {Engel}}, \bibinfo
  {author} {\bibfnamefont {H.}~\bibnamefont {Bartolf}}, \bibinfo {author}
  {\bibfnamefont {A.}~\bibnamefont {Schilling}}, \bibinfo {author}
  {\bibfnamefont {A.}~\bibnamefont {Semenov}},\ and\ \bibinfo {author}
  {\bibfnamefont {H.~W.}\ \bibnamefont {Huebers}},\ }\bibfield  {title}
  {\bibinfo {title} {Current-induced critical state in {{NbN}} thin-film
  structures},\ }\href {https://doi.org/10.1007/s10909-007-9690-5} {\bibfield
  {journal} {\bibinfo  {journal} {Journal of Low Temperature Physics}\ }\textbf
  {\bibinfo {volume} {151}},\ \bibinfo {pages} {585} (\bibinfo {year}
  {2008})}\BibitemShut {NoStop}%
\bibitem [{\citenamefont {Il'in}\ \emph {et~al.}(2005)\citenamefont {Il'in},
  \citenamefont {Siegel}, \citenamefont {Semenov}, \citenamefont {Engel},\ and\
  \citenamefont {H{\"u}bers}}]{Ilin2005}%
  \BibitemOpen
  \bibfield  {author} {\bibinfo {author} {\bibfnamefont {K.}~\bibnamefont
  {Il'in}}, \bibinfo {author} {\bibfnamefont {M.}~\bibnamefont {Siegel}},
  \bibinfo {author} {\bibfnamefont {A.}~\bibnamefont {Semenov}}, \bibinfo
  {author} {\bibfnamefont {A.}~\bibnamefont {Engel}},\ and\ \bibinfo {author}
  {\bibfnamefont {H.-W.}\ \bibnamefont {H{\"u}bers}},\ }\bibfield  {title}
  {\bibinfo {title} {Critical current of {{Nb}} and {{NbN}} thin-film
  structures: {{The}} cross-section dependence},\ }\href
  {https://doi.org/10.1002/pssc.200460811} {\bibfield  {journal} {\bibinfo
  {journal} {physica status solidi (c)}\ }\textbf {\bibinfo {volume} {2}},\
  \bibinfo {pages} {1680} (\bibinfo {year} {2005})}\BibitemShut {NoStop}%
\bibitem [{\citenamefont {Li}\ \emph {et~al.}(2019)\citenamefont {Li},
  \citenamefont {Wang}, \citenamefont {You}, \citenamefont {Wang},
  \citenamefont {Zhou}, \citenamefont {Hu}, \citenamefont {Zhang},
  \citenamefont {Liu}, \citenamefont {Yang}, \citenamefont {Zhang},
  \citenamefont {Wang},\ and\ \citenamefont {Xie}}]{Li2019b}%
  \BibitemOpen
  \bibfield  {author} {\bibinfo {author} {\bibfnamefont {H.}~\bibnamefont
  {Li}}, \bibinfo {author} {\bibfnamefont {Y.}~\bibnamefont {Wang}}, \bibinfo
  {author} {\bibfnamefont {L.}~\bibnamefont {You}}, \bibinfo {author}
  {\bibfnamefont {H.}~\bibnamefont {Wang}}, \bibinfo {author} {\bibfnamefont
  {H.}~\bibnamefont {Zhou}}, \bibinfo {author} {\bibfnamefont {P.}~\bibnamefont
  {Hu}}, \bibinfo {author} {\bibfnamefont {W.}~\bibnamefont {Zhang}}, \bibinfo
  {author} {\bibfnamefont {X.}~\bibnamefont {Liu}}, \bibinfo {author}
  {\bibfnamefont {X.}~\bibnamefont {Yang}}, \bibinfo {author} {\bibfnamefont
  {L.}~\bibnamefont {Zhang}}, \bibinfo {author} {\bibfnamefont
  {Z.}~\bibnamefont {Wang}},\ and\ \bibinfo {author} {\bibfnamefont
  {X.}~\bibnamefont {Xie}},\ }\bibfield  {title} {\bibinfo {title}
  {Supercontinuum single-photon detector using multilayer superconducting
  nanowires},\ }\href {https://doi.org/10.1364/PRJ.7.001425} {\bibfield
  {journal} {\bibinfo  {journal} {Photonics Research}\ }\textbf {\bibinfo
  {volume} {7}},\ \bibinfo {pages} {1425} (\bibinfo {year} {2019})}\BibitemShut
  {NoStop}%
\bibitem [{\citenamefont {Kerman}\ \emph {et~al.}(2006)\citenamefont {Kerman},
  \citenamefont {Dauler}, \citenamefont {Keicher}, \citenamefont {Yang},
  \citenamefont {Berggren}, \citenamefont {Gol'tsman},\ and\ \citenamefont
  {Voronov}}]{Kerman2006}%
  \BibitemOpen
  \bibfield  {author} {\bibinfo {author} {\bibfnamefont {A.~J.}\ \bibnamefont
  {Kerman}}, \bibinfo {author} {\bibfnamefont {E.~A.}\ \bibnamefont {Dauler}},
  \bibinfo {author} {\bibfnamefont {W.~E.}\ \bibnamefont {Keicher}}, \bibinfo
  {author} {\bibfnamefont {J.~K.}\ \bibnamefont {Yang}}, \bibinfo {author}
  {\bibfnamefont {K.~K.}\ \bibnamefont {Berggren}}, \bibinfo {author}
  {\bibfnamefont {G.}~\bibnamefont {Gol'tsman}},\ and\ \bibinfo {author}
  {\bibfnamefont {B.}~\bibnamefont {Voronov}},\ }\bibfield  {title} {\bibinfo
  {title} {Kinetic-inductance-limited reset time of superconducting nanowire
  photon counters},\ }\href {https://doi.org/10.1063/1.2183810} {\bibfield
  {journal} {\bibinfo  {journal} {Applied Physics Letters}\ }\textbf {\bibinfo
  {volume} {88}},\ \bibinfo {pages} {2} (\bibinfo {year} {2006})}\BibitemShut
  {NoStop}%
\bibitem [{\citenamefont {Pearl}(1964)}]{Pearl1964}%
  \BibitemOpen
  \bibfield  {author} {\bibinfo {author} {\bibfnamefont {J.}~\bibnamefont
  {Pearl}},\ }\bibfield  {title} {\bibinfo {title} {Current distribution in
  superconducting films carrying quantized fluxoids},\ }\href
  {https://doi.org/10.1063/1.1754056} {\bibfield  {journal} {\bibinfo
  {journal} {Applied Physics Letters}\ }\textbf {\bibinfo {volume} {5}},\
  \bibinfo {pages} {65} (\bibinfo {year} {1964})}\BibitemShut {NoStop}%
\bibitem [{\citenamefont {Bartolf}(2016)}]{Bartolf2016}%
  \BibitemOpen
  \bibfield  {author} {\bibinfo {author} {\bibfnamefont {H.}~\bibnamefont
  {Bartolf}},\ }\href {https://doi.org/10.1007/978-3-658-12246-1} {\emph
  {\bibinfo {title} {Fluctuation {{Mechanisms}} in {{Superconductors}}}}}\
  (\bibinfo  {publisher} {{Springer Fachmedien Wiesbaden}},\ \bibinfo {address}
  {{Wiesbaden}},\ \bibinfo {year} {2016})\BibitemShut {NoStop}%
\bibitem [{Note1()}]{Note1}%
  \BibitemOpen
  \bibinfo {note} {For a detailed derivation of \protect \cref
  {eq:kinetic-inductance-fundamental,eq:effective-magnetic-penetration-depth}
  see also the book of \protect \rev@citet [][Eqs.~(3.120), (3.137), and
  (6.67b)]{Tinkham2004}.}\BibitemShut {Stop}%
\bibitem [{\citenamefont {Alkemade}\ and\ \citenamefont {{van
  Veldhoven}}(2012)}]{Alkemade2012}%
  \BibitemOpen
  \bibfield  {author} {\bibinfo {author} {\bibfnamefont {P.~F.~A.}\
  \bibnamefont {Alkemade}}\ and\ \bibinfo {author} {\bibfnamefont
  {E.}~\bibnamefont {{van Veldhoven}}},\ }\bibfield  {title} {\bibinfo {title}
  {Deposition, {{Milling}}, and {{Etching}} with a {{Focused Helium Ion
  Beam}}},\ }in\ \href {https://doi.org/10.1007/978-3-7091-0424-8_11} {\emph
  {\bibinfo {booktitle} {Nanofabrication}}},\ \bibinfo {editor} {edited by\
  \bibinfo {editor} {\bibfnamefont {M.}~\bibnamefont {Stepanova}}\ and\
  \bibinfo {editor} {\bibfnamefont {S.}~\bibnamefont {Dew}}}\ (\bibinfo
  {publisher} {{Springer Vienna}},\ \bibinfo {address} {{Vienna}},\ \bibinfo
  {year} {2012})\ pp.\ \bibinfo {pages} {275--300}\BibitemShut {NoStop}%
\bibitem [{\citenamefont {Li}\ \emph {et~al.}(2009)\citenamefont {Li},
  \citenamefont {Martin}, \citenamefont {Anderoglu}, \citenamefont {Misra},
  \citenamefont {Shao}, \citenamefont {Wang},\ and\ \citenamefont
  {Zhang}}]{Li2009}%
  \BibitemOpen
  \bibfield  {author} {\bibinfo {author} {\bibfnamefont {N.}~\bibnamefont
  {Li}}, \bibinfo {author} {\bibfnamefont {M.~S.}\ \bibnamefont {Martin}},
  \bibinfo {author} {\bibfnamefont {O.}~\bibnamefont {Anderoglu}}, \bibinfo
  {author} {\bibfnamefont {A.}~\bibnamefont {Misra}}, \bibinfo {author}
  {\bibfnamefont {L.}~\bibnamefont {Shao}}, \bibinfo {author} {\bibfnamefont
  {H.}~\bibnamefont {Wang}},\ and\ \bibinfo {author} {\bibfnamefont
  {X.}~\bibnamefont {Zhang}},\ }\bibfield  {title} {\bibinfo {title} {He ion
  irradiation damage in {{Al}}/{{Nb}} multilayers},\ }\href
  {https://doi.org/10.1063/1.3138804} {\bibfield  {journal} {\bibinfo
  {journal} {Journal of Applied Physics}\ }\textbf {\bibinfo {volume} {105}},\
  \bibinfo {pages} {123522} (\bibinfo {year} {2009})}\BibitemShut {NoStop}%
\bibitem [{\citenamefont {Shimizu}\ \emph {et~al.}(1994)\citenamefont
  {Shimizu}, \citenamefont {Yasuda},\ and\ \citenamefont
  {Kinoshita}}]{Shimizu1994}%
  \BibitemOpen
  \bibfield  {author} {\bibinfo {author} {\bibfnamefont {J.}~\bibnamefont
  {Shimizu}}, \bibinfo {author} {\bibfnamefont {K.}~\bibnamefont {Yasuda}},\
  and\ \bibinfo {author} {\bibfnamefont {C.}~\bibnamefont {Kinoshita}},\
  }\bibfield  {title} {\bibinfo {title} {Formation process of defect clusters
  in copper and nickel under irradiation with helium ions},\ }\href
  {https://doi.org/10.1016/0022-3115(94)90057-4} {\bibfield  {journal}
  {\bibinfo  {journal} {Journal of Nuclear Materials}\ }\textbf {\bibinfo
  {volume} {212--215}},\ \bibinfo {pages} {207} (\bibinfo {year}
  {1994})}\BibitemShut {NoStop}%
\bibitem [{\citenamefont {Behrisch}\ and\ \citenamefont
  {Eckstein}(2007)}]{Behrisch2007}%
  \BibitemOpen
  \bibinfo {editor} {\bibfnamefont {R.}~\bibnamefont {Behrisch}}\ and\ \bibinfo
  {editor} {\bibfnamefont {W.}~\bibnamefont {Eckstein}},\ eds.,\ \href
  {https://doi.org/10.1007/978-3-540-44502-9} {\emph {\bibinfo {title}
  {Sputtering by {{Particle Bombardment}}}}},\ \bibinfo {edition} {1st}\ ed.,\
  \bibinfo {series} {Topics in {{Applied Physics}}}\ No.\ \bibinfo {number}
  {110}\ (\bibinfo  {publisher} {{Springer Berlin Heidelberg}},\ \bibinfo
  {address} {{Berlin, Heidelberg}},\ \bibinfo {year} {2007})\BibitemShut
  {NoStop}%
\bibitem [{\citenamefont {Ivry}\ \emph {et~al.}(2014)\citenamefont {Ivry},
  \citenamefont {Kim}, \citenamefont {Dane}, \citenamefont {De~Fazio},
  \citenamefont {McCaughan}, \citenamefont {Sunter}, \citenamefont {Zhao},\
  and\ \citenamefont {Berggren}}]{Ivry2014}%
  \BibitemOpen
  \bibfield  {author} {\bibinfo {author} {\bibfnamefont {Y.}~\bibnamefont
  {Ivry}}, \bibinfo {author} {\bibfnamefont {C.~S.}\ \bibnamefont {Kim}},
  \bibinfo {author} {\bibfnamefont {A.~E.}\ \bibnamefont {Dane}}, \bibinfo
  {author} {\bibfnamefont {D.}~\bibnamefont {De~Fazio}}, \bibinfo {author}
  {\bibfnamefont {A.~N.}\ \bibnamefont {McCaughan}}, \bibinfo {author}
  {\bibfnamefont {K.~A.}\ \bibnamefont {Sunter}}, \bibinfo {author}
  {\bibfnamefont {Q.}~\bibnamefont {Zhao}},\ and\ \bibinfo {author}
  {\bibfnamefont {K.~K.}\ \bibnamefont {Berggren}},\ }\bibfield  {title}
  {\bibinfo {title} {Universal scaling of the critical temperature for thin
  films near the superconducting-to-insulating transition},\ }\href
  {https://doi.org/10.1103/PhysRevB.90.214515} {\bibfield  {journal} {\bibinfo
  {journal} {Physical Review B}\ }\textbf {\bibinfo {volume} {90}},\ \bibinfo
  {pages} {214515} (\bibinfo {year} {2014})}\BibitemShut {NoStop}%
\bibitem [{\citenamefont {Holzman}\ and\ \citenamefont
  {Ivry}(2019)}]{Holzman2019}%
  \BibitemOpen
  \bibfield  {author} {\bibinfo {author} {\bibfnamefont {I.}~\bibnamefont
  {Holzman}}\ and\ \bibinfo {author} {\bibfnamefont {Y.}~\bibnamefont {Ivry}},\
  }\bibfield  {title} {\bibinfo {title} {Superconducting {{Nanowires}} for
  {{Single}}-{{Photon Detection}}: {{Progress}}, {{Challenges}}, and
  {{Opportunities}}},\ }\href {https://doi.org/10.1002/qute.201800058}
  {\bibfield  {journal} {\bibinfo  {journal} {Advanced Quantum Technologies}\
  }\textbf {\bibinfo {volume} {2}},\ \bibinfo {pages} {1800058} (\bibinfo
  {year} {2019})}\BibitemShut {NoStop}%
\bibitem [{\citenamefont {Haviland}\ \emph {et~al.}(1989)\citenamefont
  {Haviland}, \citenamefont {Liu},\ and\ \citenamefont
  {Goldman}}]{Haviland1989}%
  \BibitemOpen
  \bibfield  {author} {\bibinfo {author} {\bibfnamefont {D.~B.}\ \bibnamefont
  {Haviland}}, \bibinfo {author} {\bibfnamefont {Y.}~\bibnamefont {Liu}},\ and\
  \bibinfo {author} {\bibfnamefont {A.~M.}\ \bibnamefont {Goldman}},\
  }\bibfield  {title} {\bibinfo {title} {Onset of superconductivity in the
  two-dimensional limit},\ }\href {https://doi.org/10.1103/PhysRevLett.62.2180}
  {\bibfield  {journal} {\bibinfo  {journal} {Physical Review Letters}\
  }\textbf {\bibinfo {volume} {62}},\ \bibinfo {pages} {2180} (\bibinfo {year}
  {1989})}\BibitemShut {NoStop}%
\bibitem [{\citenamefont {Bezryadin}\ \emph {et~al.}(2000)\citenamefont
  {Bezryadin}, \citenamefont {Lau},\ and\ \citenamefont
  {Tinkham}}]{Bezryadin2000}%
  \BibitemOpen
  \bibfield  {author} {\bibinfo {author} {\bibfnamefont {A.}~\bibnamefont
  {Bezryadin}}, \bibinfo {author} {\bibfnamefont {C.~N.}\ \bibnamefont {Lau}},\
  and\ \bibinfo {author} {\bibfnamefont {M.}~\bibnamefont {Tinkham}},\
  }\bibfield  {title} {\bibinfo {title} {Quantum suppression of
  superconductivity in ultrathin nanowires},\ }\href
  {https://doi.org/10.1038/35010060} {\bibfield  {journal} {\bibinfo  {journal}
  {Nature}\ }\textbf {\bibinfo {volume} {404}},\ \bibinfo {pages} {971}
  (\bibinfo {year} {2000})}\BibitemShut {NoStop}%
\bibitem [{\citenamefont {Ruhtinas}\ and\ \citenamefont
  {Maasilta}(2023)}]{Ruhtinas2023}%
  \BibitemOpen
  \bibfield  {author} {\bibinfo {author} {\bibfnamefont {A.}~\bibnamefont
  {Ruhtinas}}\ and\ \bibinfo {author} {\bibfnamefont {I.~J.}\ \bibnamefont
  {Maasilta}},\ }\href@noop {} {\bibinfo {title} {Highly tunable {{NbTiN
  Josephson}} junctions fabricated with focused helium ion beam}} (\bibinfo
  {year} {2023}),\ \Eprint {https://arxiv.org/abs/2303.17348} {arxiv:2303.17348
  [cond-mat]} \BibitemShut {NoStop}%
\bibitem [{\citenamefont {Chockalingam}\ \emph {et~al.}(2008)\citenamefont
  {Chockalingam}, \citenamefont {Chand}, \citenamefont {Jesudasan},
  \citenamefont {Tripathi},\ and\ \citenamefont
  {Raychaudhuri}}]{Chockalingam2008}%
  \BibitemOpen
  \bibfield  {author} {\bibinfo {author} {\bibfnamefont {S.~P.}\ \bibnamefont
  {Chockalingam}}, \bibinfo {author} {\bibfnamefont {M.}~\bibnamefont {Chand}},
  \bibinfo {author} {\bibfnamefont {J.}~\bibnamefont {Jesudasan}}, \bibinfo
  {author} {\bibfnamefont {V.}~\bibnamefont {Tripathi}},\ and\ \bibinfo
  {author} {\bibfnamefont {P.}~\bibnamefont {Raychaudhuri}},\ }\bibfield
  {title} {\bibinfo {title} {Superconducting properties and {{Hall}} effect of
  epitaxial {{NbN}} thin films},\ }\href
  {https://doi.org/10.1103/PhysRevB.77.214503} {\bibfield  {journal} {\bibinfo
  {journal} {Physical Review B}\ }\textbf {\bibinfo {volume} {77}},\ \bibinfo
  {pages} {214503} (\bibinfo {year} {2008})}\BibitemShut {NoStop}%
\bibitem [{\citenamefont {Radebaugh}(2009)}]{Radebaugh2009}%
  \BibitemOpen
  \bibfield  {author} {\bibinfo {author} {\bibfnamefont {R.}~\bibnamefont
  {Radebaugh}},\ }\bibfield  {title} {\bibinfo {title} {Cryocoolers: The state
  of the art and recent developments},\ }\href
  {https://doi.org/10.1088/0953-8984/21/16/164219} {\bibfield  {journal}
  {\bibinfo  {journal} {Journal of Physics: Condensed Matter}\ }\textbf
  {\bibinfo {volume} {21}},\ \bibinfo {pages} {164219} (\bibinfo {year}
  {2009})}\BibitemShut {NoStop}%
\bibitem [{\citenamefont {Tinkham}(2004)}]{Tinkham2004}%
  \BibitemOpen
  \bibfield  {author} {\bibinfo {author} {\bibfnamefont {M.}~\bibnamefont
  {Tinkham}},\ }\href@noop {} {\emph {\bibinfo {title} {Introduction to
  Superconductivity}}},\ \bibinfo {edition} {2nd}\ ed.,\ Dover Books on
  Physics\ (\bibinfo  {publisher} {{Dover Publ}},\ \bibinfo {address}
  {{Mineola, NY}},\ \bibinfo {year} {2004})\BibitemShut {NoStop}%
\end{thebibliography}%

\end{document}